\documentclass[a4paper,usenatbib,times,fleqn]{mn2e}
\usepackage{natbib,graphicx,amsmath,amsfonts,amssymb,times,txfonts,pstricks,multicol}

\def\vg{\textbf{v}_{\mathrm{g}}}
\def\vd{\textbf{v}_{\mathrm{d}}}
\def\rhog{\rho_{\mathrm{g}}}
\def\rhod{\rho_{\mathrm{d}}}
\def\rhogz{\rho_{\mathrm{g},0}}
\def\rhodz{\rho_{\mathrm{d},0}}
\def\ts{t_{\mathrm{s}}}

\def\deltav{\Delta \textbf{v}}
\def\cs{c_{\mathrm{s}}}
\def\vb{\textbf{v}}

\def\rhogz{\rho_{\mathrm{g},0}}
\def\rhodz{\rho_{\mathrm{d},0}}

\def\ge{g_{\mathrm{e}}}

\def\hrhod{\hat{\rho}_{\mathrm{d}}}

\def\fg{\mathbf{f}_{\mathrm{g}}}
\def\fd{\mathbf{f}_{\mathrm{d}}}
\def\zm{z_{\mathrm{M}}}
\def\zmz{z_{\mathrm{M},0}}
\def\zdl{z_{\mathrm{d,l}}}
\def\vdl{v_{\mathrm{d,l}}}
\def\vdz{v_{\mathrm{d}}}
\def\epsd{\epsilon}
\def\epsda{\epsilon_{a}}
\def\epsdb{\epsilon_{b}}

\def\vdt{v_{\rm d}}
\def\ld{l_{\rm d}}
\def\td{t_{\rm d}}
\def\siga{\sigma_{\rm a}}
\def\sigv{\sigma_{\rm v}}
\def\deltavz{\Delta v_{0}}
\def\az{\left| a_{0} \right|}

\def\dst{\displaystyle}

\defcitealias{LP12a}{LP12a}
\defcitealias{LP12b}{LP12b}
\defcitealias{LP14a}{Paper~I}
\defcitealias{NSH86}{NSH86}

\title[Dusty gas with one fluid in SPH]{Dusty gas with one fluid in smoothed particle hydrodynamics}

\author[Laibe \& Price]{Guillaume Laibe$^{1,2}$ and Daniel J. Price$^{1}$ \\
$^{1}$Monash Centre for Astrophysics and School of Mathematical Sciences, Monash University, Clayton, Vic 3800, Australia \\
$^{2}$School of Physics and Astronomy, University of St. Andrews, North Haugh, St. Andrews, Fife KY16 9SS, UK
}
\pagerange{\pageref{firstpage}--\pageref{lastpage}} \pubyear{2014}

\begin{document}
%
%  These Macros are taken from the AAS TeX macro package version 4.0.
%  Include this file in your LaTeX source only if you are not using
%  the AAS TeX macro package and need to resolve the macro definitions
%  in the BibTeX entries returned by the ADS abstract service.
%
%  For more information on the AASTeX macro package, please see the URL
%	http://www.aas.org/publications/aastex.html
%  For more information about ADS abstract server, please see the URL
%	http://adswww.harvard.edu/ads_abstracts.html
%

% Abbreviations for journals.  The object here is to provide authors
% with convenient shorthands for the most "popular" (often-cited)
% journals; the author can use these markup tags without being concerned
% about the exact form of the journal abbreviation, or its formatting.
% It is up to the keeper of the macros to make sure the macros expand
% to the proper text.  If macro package writers agree to all use the
% same TeX command name, authors only have to remember one thing, and
% the style file will take care of editorial preferences.  This also
% applies when a single journal decides to revamp its abbreviating
% scheme, as happened with the ApJ (Abt 1991).

\def\jnl@style{\it}
%commente par Seb
\def\aaref@jnl#1{{\jnl@style#1}}
%ref remplace par aaref pour eviter conflit...

\def\aaref@jnl#1{{\jnl@style#1}}

\def\aj{\aaref@jnl{AJ}}                   % Astronomical Journal
\def\araa{\aaref@jnl{ARA\&A}}             % Annual Review of Astron and Astrophys
\def\apj{\aaref@jnl{ApJ}}                 % Astrophysical Journal
\def\apjl{\aaref@jnl{ApJ}}                % Astrophysical Journal, Letters
\def\apjs{\aaref@jnl{ApJS}}               % Astrophysical Journal, Supplement
\def\ao{\aaref@jnl{Appl.~Opt.}}           % Applied Optics
\def\apss{\aaref@jnl{Ap\&SS}}             % Astrophysics and Space Science
\def\aap{\aaref@jnl{A\&A}}                % Astronomy and Astrophysics
\def\aapr{\aaref@jnl{A\&A~Rev.}}          % Astronomy and Astrophysics Reviews
\def\aaps{\aaref@jnl{A\&AS}}              % Astronomy and Astrophysics, Supplement
\def\azh{\aaref@jnl{AZh}}                 % Astronomicheskii Zhurnal
\def\baas{\aaref@jnl{BAAS}}               % Bulletin of the AAS
\def\jrasc{\aaref@jnl{JRASC}}             % Journal of the RAS of Canada
\def\memras{\aaref@jnl{MmRAS}}            % Memoirs of the RAS
\def\mnras{\aaref@jnl{MNRAS}}             % Monthly Notices of the RAS
\def\pra{\aaref@jnl{Phys.~Rev.~A}}        % Physical Review A: General Physics
\def\prb{\aaref@jnl{Phys.~Rev.~B}}        % Physical Review B: Solid State
\def\prc{\aaref@jnl{Phys.~Rev.~C}}        % Physical Review C
\def\prd{\aaref@jnl{Phys.~Rev.~D}}        % Physical Review D
\def\pre{\aaref@jnl{Phys.~Rev.~E}}        % Physical Review E
\def\prl{\aaref@jnl{Phys.~Rev.~Lett.}}    % Physical Review Letters
\def\pasp{\aaref@jnl{PASP}}               % Publications of the ASP
\def\pasj{\aaref@jnl{PASJ}}               % Publications of the ASJ
\def\qjras{\aaref@jnl{QJRAS}}             % Quarterly Journal of the RAS
\def\skytel{\aaref@jnl{S\&T}}             % Sky and Telescope
\def\solphys{\aaref@jnl{Sol.~Phys.}}      % Solar Physics
\def\sovast{\aaref@jnl{Soviet~Ast.}}      % Soviet Astronomy
\def\ssr{\aaref@jnl{Space~Sci.~Rev.}}     % Space Science Reviews
\def\zap{\aaref@jnl{ZAp}}                 % Zeitschrift fuer Astrophysik
\def\nat{\aaref@jnl{Nature}}              % Nature
\def\iaucirc{\aaref@jnl{IAU~Circ.}}       % IAU Cirulars
\def\aplett{\aaref@jnl{Astrophys.~Lett.}} % Astrophysics Letters
\def\apspr{\aaref@jnl{Astrophys.~Space~Phys.~Res.}}
                % Astrophysics Space Physics Research
\def\bain{\aaref@jnl{Bull.~Astron.~Inst.~Netherlands}} 
                % Bulletin Astronomical Institute of the Netherlands
\def\fcp{\aaref@jnl{Fund.~Cosmic~Phys.}}  % Fundamental Cosmic Physics
\def\gca{\aaref@jnl{Geochim.~Cosmochim.~Acta}}   % Geochimica Cosmochimica Acta
\def\grl{\aaref@jnl{Geophys.~Res.~Lett.}} % Geophysics Research Letters
\def\jcp{\aaref@jnl{J.~Chem.~Phys.}}      % Journal of Chemical Physics
\def\jgr{\aaref@jnl{J.~Geophys.~Res.}}    % Journal of Geophysics Research
\def\jqsrt{\aaref@jnl{J.~Quant.~Spec.~Radiat.~Transf.}}
                % Journal of Quantitiative Spectroscopy and Radiative Transfer
\def\memsai{\aaref@jnl{Mem.~Soc.~Astron.~Italiana}}
                % Mem. Societa Astronomica Italiana
\def\nphysa{\aaref@jnl{Nucl.~Phys.~A}}   % Nuclear Physics A
\def\physrep{\aaref@jnl{Phys.~Rep.}}   % Physics Reports
\def\physscr{\aaref@jnl{Phys.~Scr}}   % Physica Scripta
\def\planss{\aaref@jnl{Planet.~Space~Sci.}}   % Planetary Space Science
\def\procspie{\aaref@jnl{Proc.~SPIE}}   % Proceedings of the SPIE

\let\astap=\aap
\let\apjlett=\apjl
\let\apjsupp=\apjs
\let\applopt=\ao

\label{firstpage}
\bibliographystyle{mn2e}
\maketitle

\begin{abstract}
In a companion paper we have shown how the equations describing gas and dust as two fluids coupled by a drag term can be re-formulated to describe the system as a single fluid mixture. Here we present a numerical implementation of the one-fluid dusty gas algorithm using Smoothed Particle Hydrodynamics (SPH).

The algorithm preserves the conservation properties of the SPH formalism. In particular, the total gas and dust mass, momentum, angular momentum and energy are all exactly conserved. Shock viscosity and conductivity terms are generalised to handle the two-phase mixture accordingly. The algorithm is benchmarked against a comprehensive suit of problems: \textsc{dustybox}, \textsc{dustywave}, \textsc{dustyshock} and \textsc{dustyoscill}, each of them addressing different properties of the method. We compare the performance of the one-fluid algorithm to the standard two-fluid approach.

The one-fluid algorithm is found to solve both of the fundamental limitations of the two-fluid algorithm: it is no longer possible to concentrate dust below the resolution of the gas (they have the same resolution by definition), and the spatial resolution criterion $h < c_{\rm s}t_{\rm s}$, required in two-fluid codes to avoid over-damping of kinetic energy, is unnecessary. Implicit time stepping is straightforward. As a result, the algorithm is up to ten billion times more efficient for 3D simulations of small grains. Additional benefits include the use of half as many particles, a single kernel and fewer SPH interpolations. The only limitation is that it does not capture multi-streaming of dust in the limit of zero coupling, suggesting that in this case a hybrid approach may be required.
\end{abstract}


%----------------------------------------------------------------------------------------------------------------
\section{Introduction}
\label{sec:intro}

 Gas and dust are the two primary constituents of cold astrophysical systems. Observations of such systems mainly probe the dust phase, with gas properties deduced from the dust distribution assuming a constant $\simeq 10^{-2}$ dust-to-gas ratio, typical of the interstellar medium. However, the evolution of the gas and dust phases may be quite different, particularly during planet formation. Once the homogeneous mixture assumption breaks down, numerical simulations of the full gas and dust mixture become the only reliable way to predict the properties of the dust phase. Indeed, with the exception of a few problems, the evolution equations for the gas and dust mixture are not tractable analytically. The dust phase is usually treated as a continuous pressureless fluid, an approximation which captures the physics of the system (especially for small grains) but enormously simplifies its description \citep{Garaud2004}. In particular, the fluid equations are easier to handle compared to integrating the motion of individual grains. 
 
Several numerical methods, both grid-based or Lagrangian, exist to discretise the equations of evolution of a gas and dust mixture. In this paper, we focus on the Smoothed Particle Hydrodynamics (SPH) approach, a fully conservative Lagrangian numerical method (e.g. \citealt{Monaghan2012,price12}). SPH is well suited to problems with large density contrasts, free boundaries and complex geometries.  Moreover, its Lagrangian nature couples perfectly with the Lagrangian nature of the dust dynamics, making SPH particularly useful for dust-gas mixtures in astrophysics.  \citet{Monaghan1995} performed the foundational work on a two-fluid SPH gas and dust algorithm, where drag terms were computed by projecting the differential velocity between the particles onto their line of sight to preserve the exact conservation of angular momentum. \citet{Monaghandust97} improved the method with an implicit timestepping method designed to handle strong drag regimes. This SPH algorithm has been used mostly in the context of planet formation, where grains differentiate strongly from the gas phase \citep{Maddison2003,Rice04,BF2005,Pinte2007,Maddison2007,Fouchet2007,Laibe2008,Fouchet2010}.
 
In \citet{LP12a,LP12b} we highlighted and addressed a number of issues with the existing two fluid SPH algorithm. These included the generalised SPH density estimate in multi-fluid systems, the consistent treatment of variable smoothing length terms and finite particle size, timestep stability, implicit integration, the treatment of non-linear drag regimes, thermal coupling terms and the choice of kernel and smoothing length used in the drag operator. A rigorous mathematical analysis of the algorithm was performed and it was validated against a series of tests, including the analytic solutions recently derived in \citet{LP11}. The algorithm has been used to predict the filtering and concentration of solids by an existing planet embedded in a protoplanetary disc \citep{Ayliffe2012}.

Although this improved two-fluid algorithm was found to perform satisfactorily in a number of astrophysical contexts, some important limitations remained \citep{LP12a,LP12b}. Some of these related to the specific formulation in SPH --- for example, modelling the two fluids as distinct sets of particles meant that interpolation between the two fluids is required. This made it difficult to derive a fast and accurate implicit integration, necessary when the drag is strong, while preserving the exact conservation of angular momentum that is one of the main strengths of the SPH approach \citep{LP12b}. 

More fundamentally, the two-fluid algorithm was found to have two major limitations not specific to SPH \citep{LP12a}. The first limitation was that both infinite spatial and infinite temporal resolution are required in the rather obvious limit of infinite drag ($\ts \to 0$, where $\ts$ is the stopping time), where the fluids become perfectly coupled. The timestep restriction $\Delta t < \ts$ was already well known and can be handled using implicit timestepping methods, as described in \citep{LP12b}. However, we also found a \emph{spatial} resolution criterion $\Delta x \lesssim \cs \ts$, where $\cs$ is the gas sound speed, in order to prevent artificial separation and thus overdamping of the two-fluid mixture. Since $\ts$ is very small for strong drag (small grains), this implies a prohibitive computational cost, especially for three dimensional simulations the additional expense may be a factor of several thousand or even millions for micron to millimetre-size grains in protoplanetary discs).  The second limitation was that the total resolution of the simulation is fixed by the more poorly resolved phase. As a result, the structures produced when the dust over concentrates with respect to the gas were found to be strongly resolution-dependent and hence artificial. This prevents the study of systems in which dust is expected to concentrate by orders of magnitude, such as planet formation.
 
 These limitations, which we see as fundamental, have been ignored in almost all gas and dust simulations to date. They are intrinsic
to the two-fluid description of the gas and dust mixture. We therefore radically changed our line of attack on the problem. In a companion paper (\citealt{LP14a}, hereafter \citetalias{LP14a}) we have proposed a formulation where the gas and dust are described as a \textit{single fluid made of two different phases}, referred to as \textit{the mixture}, instead of two fluids interacting with each other. The equations describing the evolution of the mixture were derived. The mixture's mass is the sum of the masses of the two phases and is advected at the barycentric velocity of the two phases. The differential velocity between the two phases and the dust-to-gas ratio, or equivalently the dust fraction, appear naturally as \textit{internal} properties of the mixture. Treating the dust and gas system as a single fluid naturally solves the two fundamental difficulties described above: In the strong drag limit the equations reduce to the usual fluid equations with a modified sound speed. Secondly. the one-fluid approach implies one resolution length for the system as a whole, meaning that no resolution criterion is required. The one-fluid equations were also found to provide a better physical insight into the behaviour of the mixture compared to their two-fluid counterparts.

In this paper, we derive and test an SPH algorithm for dust-gas mixtures based on the single fluid description. This algorithm is conservative \textit{by construction}: it conserves the mass, the momentum, angular momentum and the energy of the mixture exactly, as well as the specific mass of each phase taken individually. Furthermore, each SPH particle carries information regarding both phases, removing the need for additional interpolation and/or complicated timestepping schemes. 

 In Section~\ref{sec:onefluid} we summarise the relevant parts of the one-fluid formulation from 
\citetalias{LP14a}. In Sect.~\ref{sec:SPH} we derive the SPH versions of these equations. We also discuss timestepping and the necessary modifications to the SPH shock capturing terms. The algorithm is validated against a suite of tests in Sect.~\ref{sec:tests}, comparing the results against analytic solutions and to those obtained with our previous two-fluid algorithm. We summarise the advantages of the one-fluid approach in Sect.~\ref{sec:conclu}, and also point out some limitations of this current version of the algorithm in the limit of weak/zero drag.

%=======================================================================================================
\section{Dusty gas with one fluid}
\label{sec:onefluid}

\subsection{Two fluid dust and gas mixtures}
\label{sec:evoleqtwo}

The equations for the conservation of density and momentum are given by \citepalias{LP14a}:
\begin{eqnarray}
\frac{\partial \rhog}{\partial t} + \nabla . \left ( \rhog \vg \right) & = & 0 \label{eq:mass_gas},\\
\frac{\partial \rhod}{\partial t} + \nabla . \left ( \rhod \vd \right) & = & 0 \label{eq:mass_dust}, \\
\rhog \left( \frac{\partial \vg}{\partial t} + \vg . \nabla \vg \right)  & = & \rhog \fg +  K  (\vd - \vg) +  \rhog \textbf{f} \label{eq:momentum_gas},\\[1em]
\rhod \left( \frac{\partial \vd}{\partial t} + \vd . \nabla \vd \right) & = &\rhod \fd  -  K  (\vd - \vg) + \rhod \textbf{f}  \label{eq:momentum_dust},
\end{eqnarray}
where $K$ is the drag coefficient which is a function of the local gas and dust parameters, as well as the differential velocity between the fluids. $\fg$ and $\fd$ denote the forces that are specific to the gas and the dust phases respectively. The gas force, $\fg$, contains at least the pressure gradient term $- \phantom{.}\nabla P_{\rm g} / \rhog$ (when the buoyancy term coming from the dust is negligible), but can include additional terms such as viscosity. In most of astrophysical situations, $\fd = 0$, but in general, this term may contain the gas buoyancy and/or the intrinsic dust pressure and viscosity. The internal energy equation is given by
\begin{equation}
\frac{\partial u}{\partial t} + (\vg . \nabla) u =  -\frac{P_{\mathrm{g}}}{\rhog} (\nabla . \vg) + \frac{K}{\rhog} (\vd - \vg)^{2} + \Lambda. \label{eq:newu}
\end{equation}
The equation set is closed by an equation of state. Unless otherwise specified, for the tests in this paper we adopt the ideal gas equation of state given by
\begin{equation}
P_{\rm g} = (\gamma - 1)\rho_{\rm g} u.
\end{equation}

\subsection{Dusty gas with one fluid}
\label{sec:evoleqone}

In a companion paper \citep[hereafter Paper~I]{LP14a}, we have shown that Eqs.~\ref{eq:mass_gas} -- \ref{eq:momentum_dust} can be reformulated as a single fluid, moving with the barycentric velocity,
\begin{eqnarray}
\vb & \equiv & \frac{\rhog \vg + \rhod \vd}{\rhog + \rhod} \label{eq:def_vb} .
\end{eqnarray}
The differential velocity between the two phases, $\deltav$, is defined according to
\begin{eqnarray}
\Delta {\bf v} & \equiv & \vd -\vg  \label{eq:def_deltav}.
\end{eqnarray}
The total density $\rho \equiv \rhog + \rhod$ and the dust fraction $\epsd = \rhod / \rho$ naturally replace the gas and the dust densities involved in the two fluids formalism. Using these variables, Eqs.~\ref{eq:mass_gas} -- \ref{eq:momentum_dust} become \citepalias{LP14a}
\begin{eqnarray}
\frac{{\rm d} \rho}{{\rm d} t}& = & - \rho (\nabla . \vb), \label{eq:genmass_rho} \\
\frac{{\rm d} \epsd}{{\rm d} t}& = & -\frac{1}{\rho} \nabla \cdot \left[ \epsd\left(1 - \epsd \right) \rho \deltav \right] \label{eq:gendtgevol} , \\
\frac{{\rm d} \vb}{{\rm d} t} & = & \left(1 - \epsd \right)\mathbf{f}_{\rm g} + \epsd \mathbf{f}_{\rm d} - \frac{1}{\rho}\nabla\cdot \left[ \epsd\left(1 - \epsd \right) \rho \deltav \deltav \right] + \mathbf{f}\label{eq:genmomentum_bary},\\[1em]
\frac{{\rm d} \deltav}{{\rm d} t}  & = &  - \frac{\deltav}{\ts} + ({\bf f}_{\rm d} - {\bf f}_{\rm g}) - (\deltav \cdot \nabla) \vb + \frac{1}{2}\nabla \left[ \left(2 \epsd - 1 \right) \deltav ^{2} \right], \label{eq:genmomentum_deltav}
\end{eqnarray}
where the comoving derivative refers to a particle moving with the barycentric velocity $\vb$, i.e.
\begin{equation}
\frac{{\rm d}}{{\rm d}t} \equiv \frac{\partial}{\partial t} + (\vb. \nabla) ,
\end{equation}
and the stopping time $\ts$ is given by
\begin{equation}
\ts = \frac{ \epsd\left(1 - \epsd \right) \rho}{K } .
\label{eq:def_ts}
\end{equation}
Eq.~\ref{eq:newu} becomes
\begin{equation}
\frac{{\rm d} u}{{\rm d} t} =  -\frac{ P_{\mathrm{g}}}{\left(1 - \epsd \right) \rho} \nabla . \left( \vb - \epsd \deltav \right) +   \epsd \left( \deltav . \nabla\right) u  + \epsd  \frac{\deltav^{2}}{\ts}. \label{eq:newusingle}
\end{equation}

%% EQUATIONS WITH THE DUST TO GAS RATIO
%\begin{eqnarray}
%\frac{{\rm d} \rho}{{\rm d} t}& = & - \rho (\nabla . \vb), \label{eq:genmass_rho} \\
%\frac{{\rm d}}{{\rm d} t} \left(\frac{\rhod}{\rhog} \right) & = & -\frac{\rho}{\rhog^{2}} \nabla \cdot \left(\frac{\rhog \rhod}{\rho} \deltav \right) \label{eq:gendtgevol} , \\
%\frac{{\rm d} \vb}{{\rm d} t} & = & \frac{\rhog}{\rho} \mathbf{f}_{\rm g} + \frac{\rhod}{\rho} \mathbf{f}_{\rm d} - \frac{1}{\rho}\nabla\cdot \left(\frac{\rhog \rhod}{\rho} \deltav \deltav \right) + \mathbf{f}\label{eq:genmomentum_bary},\\[1em]
%\frac{{\rm d} \deltav}{{\rm d} t}  & = &  - \frac{\deltav}{\ts} + ({\bf f}_{\rm d} - {\bf f}_{\rm g}) - (\deltav \cdot \nabla) \vb + \frac{1}{2}\nabla \left( \frac{\rhod - \rhog}{\rhod + \rhog} \deltav ^{2} \right), \label{eq:genmomentum_deltav}
%\end{eqnarray}
%where the comoving derivative refers to a particle moving with the barycentric velocity $\vb$, i.e.
%\begin{equation}
%\frac{{\rm d}}{{\rm d}t} \equiv \frac{\partial}{\partial t} + (\vb. \nabla).
%\end{equation}
%Eq.~\ref{eq:newu} becomes
%\begin{equation}
%\frac{{\rm d} u}{{\rm d} t} =  -\frac{P_{\mathrm{g}}}{\rhog} (\nabla . \vg) +   \frac{\rhod}{\rho}\left( \deltav . \nabla\right) u  + \frac{\rhod}{\rho} \frac{\deltav^{2}}{\ts}. \label{eq:newusingle}
%\end{equation}

\section{SPH formalism}
\label{sec:SPH}
 We solve Eqs.~\ref{eq:genmass_rho}--\ref{eq:newusingle} using Smoothed Particle Hydrodynamics \citep{monaghan05,price12}. The key point of difference to our earlier formulation of dusty gas in SPH \citep{LP12a,LP12b} is that rather than considering separate `gas' and `dust' particles as two interpenetrating fluids, here we consider only \emph{one} set of particles representing the mixture. These particles are of fixed mass $m$ and are moved with the barycentric velocity, i.e.
\begin{equation}
\frac{{\rm d}{\bf x}_{a}}{{\rm d}t} = {\bf v}_{a}, \label{eq:dxdt}
\end{equation}
where here and throughout the paper we use $a$ and $b$ as labels referring to the particle index.

\subsection{Conserved quantities}
\label{sec:conserved}
The SPH formulation is constrained by the conservation properties of the mixture (c.f. \citetalias{LP14a}). The total mass $M$, linear momentum $\mathbf{P}$ and energy $E$ of the mixture are given in SPH form by
\begin{eqnarray}
M & \equiv & \int \rho {\rm d}V = \sum_{a} m_{a} \label{eq:newm},\\
\mathbf{P} & \equiv & \int \rho \vb {\rm d}V = \sum_{a} m_{a} {\bf v}_{a} \label{eq:newp},\\
E & \equiv & \int \left( \frac{1}{2}\rho \vb^{2} + \frac{1}{2}\epsd\left(1 - \epsd \right) \rho \deltav^{2} + \left(1 - \epsd \right)\rho u  \right) {\rm d}V  , \nonumber \\
& = & \sum_{a} m_{a}  \left( \frac{1}{2}\vb_{a}^{2} + \frac{1}{2}\epsda \left(1 - \epsda \right) \deltav^{2}_{a} +  \left(1 - \epsda \right) u_{a}  \right) \label{eq:newe}
\end{eqnarray}
where to convert the integrals to SPH, we have rewritten the integral as a sum, with the mass element $\rho {\rm d}V$ replaced by the particle mass $m$. Similarly, the total mass of the gas and dust are independently conserved,
\begin{eqnarray}
M_{\rm g} & \equiv & \int \rhog {\rm d}V = \int \left(1 - \epsd \right) \rho {\rm d}V = \sum_{a} m_{a} \left(1 - \epsda \right) ,\label{eq:mg} \\
M_{\rm d} & \equiv & \int \rhod {\rm d}V = \int \epsd \rho {\rm d}V  = \sum_{a} m_{a} \epsda  \label{eq:md} .
\end{eqnarray}

%WITH THE DTG
%\begin{eqnarray}
%M & \equiv & \int \rho {\rm d}V = \sum_{a} m_{a} \label{eq:newm},\\
%\mathbf{P} & \equiv & \int \rho \vb {\rm d}V = \sum_{a} m_{a} {\bf v}_{a} \label{eq:newp},\\
%E & \equiv & \int \left( \frac{1}{2}\rho \vb^{2} + \frac{1}{2}\frac{\rhog \rhod}{\rho} \deltav^{2} + \rhog u  \right) {\rm d}V  , \nonumber \\
%& = & \sum_{a} m_{a}  \left( \frac{1}{2}\vb_{a}^{2} + \frac{1}{2}\frac{\rhog \rhod}{\rho_{a}^{2}} \deltav^{2}_{a} + \frac{\rhog}{\rho_{a}} u_{a}  \right) \label{eq:newe}
%\end{eqnarray}
%where to convert the integrals to SPH, we rewrite the integral as a sum, with the mass element $\rho {\rm d}V$ replaced by the particle mass $m$. Similarly, the total mass of the gas and dust are independently conserved,
%\begin{eqnarray}
%M_{\rm g} & \equiv & \int \rhog {\rm d}V = \int \frac{\rho}{1 + \rhod/\rhog} {\rm d}V  = \sum_{a} m_{a} \frac{1}{1 + \rhod/\rhog}\label{eq:mg}, \\
%M_{\rm d} & \equiv & \int \rhod {\rm d}V = \int \frac{\rho (\rhod/\rhog)}{1 + \rhod/\rhog} {\rm d}V = \sum_{a} m_{a} \frac{\rhod/\rhog}{1 + \rhod/\rhog} \label{eq:md}.
%\end{eqnarray}
As we will show below, it is possible to exactly and simultaneously conserve all of the above properties in our SPH formulation. This means in practice that the conservation properties are determined entirely by the timestepping scheme.

\subsection{Densities}
\label{sec:densities}
Eq.~\ref{eq:genmass_rho} is solved in the usual manner using the SPH density sum,
\begin{equation}
\rho_{a} = \sum_{b} m_{b} W_{ab} (h_{a}),
\label{eq:rhosum}
\end{equation}
where $W_{ab}(h) \equiv W(\vert {\bf r}_{a} - {\bf r}_{b} \vert, h)$ is the smoothing kernel and we follow the usual approach where the smoothing length is adapted according to the mean local particle spacing using
\begin{equation}
h_{a} = \eta \left(\frac{m_{a}}{\rho_{a}}\right)^{1/\nu},
\label{eq:hfac}
\end{equation}
where $\nu$ is the number of spatial dimensions and $\eta = 1.2$ is a proportionality factor that determines the mean neighbour number \citep{price12}. As is usual practice we solve Eqs.~\ref{eq:rhosum} and \ref{eq:hfac} simultaneously using an iterative Newton-Raphson method \citep{pricemonaghan04b,pricemonaghan07}.

That Eq.~\ref{eq:rhosum} is a time-independent solution to Eq.~\ref{eq:genmass_rho} follows directly from Eq.~\ref{eq:dxdt}. Taking the time derivative of Eq.~\ref{eq:rhosum} gives
\begin{equation}
\frac{{\rm d}\rho_{a}}{{\rm d} t} = \frac{1}{\Omega_{a}} \sum_{b} m_{b} \left({\bf v}_{a} - {\bf v}_{b}\right) \cdot \nabla W_{ab} (h_{a}),
\label{eq:drhodtSPH}
\end{equation}
where
\begin{equation}
\Omega_{a} = 1 - \frac{\partial h_{a}}{\partial\rho_{a}} \sum_{b} m_{b} \frac{\partial W_{ab}(h_{a})}{h_{a}}.
\end{equation}
Eq.~\ref{eq:drhodtSPH} is an SPH representation of 
\begin{equation}
\frac{{\rm d}\rho}{{\rm d} t} = \vb.\nabla\rho - \nabla. (\rho {\bf v}) = -\rho (\nabla.{\bf v}),
\end{equation}
i.e. Eq.~\ref{eq:genmass_rho}. The total mass of the mixture (Eq.~\ref{eq:newm}) is trivially conserved since the mass of each mixture particle is constant.

We note that no special kernels are required for the one fluid mixture, unlike the two-fluid case \citep{LP12a}, since there is no interpolation required between the two phases --- the densities and velocities of the two phases of the mixture are both carried by the same particle. Hence $W_{ab}$ above refers to the usual bell-shaped kernel, such as the $M_{4}$ cubic spline or the $M_{6}$ quintic spline \citep[e.g.][]{monaghan92}. We use the $M_{4}$ cubic spline in this paper.

\subsection{Dust-to-gas ratio evolution}
\label{sec:dtg}
 The SPH formulation of Eq.~\ref{eq:gendtgevol} is constrained by the requirement that the total mass of gas and the total mass of dust are independently conserved. From Eq.~\ref{eq:mg} the constraint on the gas and the dust masses corresponds to
\begin{equation}
\frac{{\rm d}M_{\rm d}}{{\rm d}t} = - \frac{{\rm d}M_{\rm g}}{{\rm d}t} =  \sum_{a} m_{a}  \frac{{\rm d} \epsda }{{\rm d} t} = 0.
\label{eq:dmgasdt}
\end{equation} 

%%WITH DTG 
%\begin{equation}
%\frac{{\rm d}M_{\rm g}}{{\rm d}t} = -\sum_{a} m_{a} \frac{1}{(1 + \epsilon_{a})^{2}}\frac{{\rm d}}{{\rm d} t} \left(\frac{\rhod}{\rhog} \right) = 0.
%\label{eq:dmgasdt}
%\end{equation}
%TO GET RID OF 
%Similarly, the constraint on the total dust mass is given by
%\begin{equation}
%\frac{{\rm d}M_{\rm d}}{{\rm d}t} = \sum_{a} m_{a} \frac{1}{(1 + \epsilon_{a})^{2}}\frac{{\rm d}}{{\rm d} t} \left(\frac{\rhod}{\rhog} \right) = 0.
%\label{eq:dmdustdt}
%\end{equation}

Substituting the continuum expression for ${\rm d}\epsd/{\rm d}t$ from Eq.~\ref{eq:gendtgevol} we find a constraint on the derivative operator of the form
\begin{equation}
\sum_{a} m_{a} \frac{1}{\rho_{a}} \nabla \cdot \left[\epsd \left(1 - \epsd \right)\rho \deltav \right] = 0.
\end{equation}

%WITH DTG
%\begin{equation}
%\sum_{a} m_{a} \frac{1}{\rho_{a}} \nabla \cdot \left(\frac{\rhog \rhod}{\rho} \deltav \right) = 0.
%\end{equation}
This constraint is satisfied using the symmetric SPH divergence operator \citep[e.g.][]{price12,triccoprice12}. Hence, the SPH expression for Eq.~\ref{eq:gendtgevol} that satisfies the conservation laws is given by
\begin{align}
\frac{{\rm d}\epsda}{{\rm d} t} = - \sum_{b} m_{b} & \left[ \frac{\epsda \left(1 - \epsda \right)}{\Omega_{a} \rho_{a}}  \deltav_{a}\cdot\nabla_{a} W_{ab} (h_{a})\right. \nonumber \\
 & + \left. \frac{\epsdb \left(1 - \epsdb \right)}{\Omega_{b} \rho_{b}}\deltav_{b}\cdot \nabla_{a} W_{ab} (h_{b}) \right]. \label{eq:dusttogassph}
\end{align}
%WITH DTG
%\begin{equation}
%\frac{{\rm d}}{{\rm d} t} \left(\frac{\rhod}{\rhog} \right) = -\frac{\rho_{a}^{2}}{\rho_{{\rm g},a}^{2}} \sum_{b} m_{b} \left[ \frac{\rhog^{a}\rhod^{a} \deltav_{a}}{\Omega_{a} \rho_{a}^{3}} .\nabla W_{ab} (h_{a}) + \frac{\rhog^{b}\rhod^{b} \deltav_{b}}{\Omega_{b} \rho_{b}^{3}}\cdot \nabla W_{ab} (h_{b}) \right]. \label{eq:dusttogassph}
%\end{equation}
Substitution of this expression in Eq.~\ref{eq:dmgasdt}, and using the antisymmetry of the kernel gradient $\nabla_{a} W_{ab} = -\nabla_{b} W_{ba}$, gives zero in the double summation. Hence, the total mass of each species is exactly conserved.

\subsection{Momentum equation}
\label{sec:momentum}
The requirement of momentum conservation constrains the formulation of Eq.~\ref{eq:genmomentum_bary}. This constraint is readily satisfied by a straightforward generalisation of the usual SPH acceleration equation for a single gaseous fluid. Here we assume that ${\bf f}_{d} = 0$ and that the forces on the gas are given by the usual hydrodynamic forces
\begin{equation}
{\bf f}_{g} = -\frac{\nabla P_{g}}{\left(1 - \epsd \right)\rho} + {\bf f}_{\rm g,visc},
\end{equation}
%WITH DTG
%\begin{equation}
%{\bf f}_{g} = -\frac{\nabla P_{g}}{\rhog} + {\bf f}_{\rm g,visc},
%\end{equation}
giving
\begin{equation}
\frac{{\rm d} \vb}{{\rm d} t} = -\frac{\nabla P_{g}}{\rho} - \frac{1}{\rho}\nabla\cdot \left[\epsd \left(1 - \epsd \right)\rho \deltav \deltav \right] + \left(1 - \epsd \right){\bf f}_{\rm g,visc} + {\bf f},
\end{equation}

%WITH DTG
%\begin{equation}
%\frac{{\rm d} \vb}{{\rm d} t} = -\frac{\nabla P_{g}}{\rho} - \frac{1}{\rho}\nabla\cdot \left(\frac{\rhog \rhod}{\rho \deltav \deltav \right) + \frac{\rhog}{\rho}{\bf f}_{\rm g,visc} + {\bf f},
%\end{equation}
where ${\bf f}_{\rm visc}$ is an (as yet unspecified) viscosity term. In this case, the SPH acceleration equation in momentum-conserving form is given by
\begin{align}
\frac{{\rm d} \vb_{a}}{{\rm d} t} = & -\sum_{b} m_{b} \left[ \frac{P_{a}}{\Omega_{a} \rho_{a}^{2}} \nabla W_{ab}(h_{a}) + \frac{P_{b}}{\Omega_{b} \rho_{b}^{2}} \nabla W_{ab}(h_{b}) \right] \nonumber \\
& -\sum_{b} m_{b} \left[ \frac{\epsda\left(1 - \epsda \right) \deltav_{a}}{\Omega_{a} \rho_{a}} \deltav_{a}.\nabla W_{ab}(h_{a}) \right. \\
&  \phantom{\sum_{b} m_{b}[} + \left.\frac{\epsdb\left(1 - \epsdb \right)\deltav_{b}}{\Omega_{b} \rho_{b}} \deltav_{b}.\nabla W_{ab}(h_{b}) \right] \nonumber \\
& +  (1 - \epsd_{a}){\bf f}_{\rm g,visc} + {\bf f}_{a}.
\label{eq:sphmom}
\end{align}
%WITH DTG
%\begin{align}
%\frac{{\rm d} \vb_{a}}{{\rm d} t} = & -\sum_{b} m_{b} \left[ \frac{P_{a}}{\Omega_{a} \rho_{a}^{2}} \nabla W_{ab}(h_{a}) + \frac{P_{b}}{\Omega_{b} \rho_{b}^{2}} \nabla W_{ab}(h_{b}) \right] \nonumber \\
%& -\sum_{b} m_{b} \left[ \frac{\rhog\rhod \deltav_{a}}{\Omega_{a} \rho_{a}^{3}} \deltav_{a}.\nabla W_{ab}(h_{a}) + \frac{\rhog\rhod\deltav_{b}}{\Omega_{b} \rho_{b}^{3}} \deltav_{b}.\nabla W_{ab}(h_{b}) \right] \nonumber \\
%& + \frac{\rhog^{a}}{\rho_{a}}{\bf f}_{\rm g,visc} + {\bf f}_{a}.
%\label{eq:sphmom}
%\end{align}
It may be straightforwardly verified that the momentum is exactly conserved, since
\begin{equation}
\sum_{a} m_{a} \frac{{\rm d} \vb_{a}}{{\rm d} t} = 0.
\label{eq:momcons}
\end{equation}
Using the formulation of Eq.~\ref{eq:sphmom} above also has the advantage that the SPH equations reduce exactly to the usual SPH equations for gas dynamics in the limit where there is no dust.
%
%Interestingly, Eq.~\ref{eq:sphmom} can be derived from a Lagrangian that generalises the usual SPH Lagrangian for a gas. The total energy of the mixture is
%\begin{equation}
%E = E_{\rm c} + E_{\rm p} =  \frac{1}{2} m \vb^{2} + U(S,V,m\deltav) ,
%\end{equation}
%%
%where
%%
%\begin{equation}
% U(S,V,m\deltav) = U_{\rm g}(S,V) +\epsd \left(1 - \epsd \right) \deltav \times \left( m \deltav \right).
%\end{equation}
%%
%Thus, the total differential of the internal energy per unit mass is:
%%
%\begin{equation}
%\mathrm{d} u = T \mathrm{d} S + \frac{P}{\rho^{2}} \mathrm{d} \rho + \left\lbrace \epsd \left(1 - \epsd \right) \deltav \right\rbrace\mathrm{d}\left( \rho \deltav \right) .
%\end{equation}
%%
%In SPH,
%%
%\begin{equation}
%\left(  \rho \deltav \right)_{a} = \sum_{b} m_{a} \deltav_{a} W_{ab} 
%\end{equation}
%%
%and Eq.~\ref{eq:sphmom} comes straight from a variational principle, using the usual SPH procedure. As a remark, the term $\rho \deltav$ is an extensive thermodynamical variable of the system. To achieve thermal equilibrium, the mixture generates a flux which diffuses its intensive conjugates variable $ \epsd \left(1 - \epsd \right) \deltav$. The max of $ \epsd \left(1 - \epsd \right)$ is obtained for $\epsd = 1/2$, i.e. when $\rhog = \rhod$ (homogeneous mixture). Such an homogenisation may be achieved by microscopic collisions or by dusty turbulence.

\subsection{Differential velocity and energy equations}
\label{sec:differential velocities}
The SPH form of the remaining two equations \ref{eq:genmomentum_deltav} and \ref{eq:newusingle} are determined by the requirement of energy conservation. Taking the time derivative of Eq.~\ref{eq:newe} we find that energy conservation is expressed by the condition
\begin{align}
\sum_{a} m_{a} \Bigg\{ &{\bf v}_{a}. \frac{{\rm d}{\bf v}_{a}}{{\rm d}t} +
 \epsda \left(1 - \epsda \right)  \deltav_{a}\cdot \frac{{\rm d}{\deltav}_{a}}{{\rm d}t} \nonumber \\
& +\left[ \left(1 - 2 \epsda \right) \frac{\deltav_{a}^{2}}{2} - u_{a}\right] \frac{{\rm d}\epsilon_{a}}{{\rm d}t}
 + \left(1 - \epsda \right) \frac{{\rm d}u_{a}}{{\rm d}t}
\Bigg\} = 0.
\label{eq:dedt}
\end{align}

%WITH DTG
%\begin{align}
%\sum_{a} m_{a} \Bigg\{ &{\bf v}_{a}. \frac{{\rm d}{\bf v}_{a}}{{\rm d}t} +
% \frac{\rhog\rhod}{\rho_{a}^{2}} \deltav_{a}\cdot \frac{{\rm d}{\deltav}_{a}}{{\rm d}t} \nonumber \\
%& + \frac{\rhog^{2}}{\rho_{a}^{2}}\left[ \frac{1 - \epsilon_{a}}{1 + \epsilon_{a}} \frac12\deltav_{a}^{2} - u_{a}\right] \frac{{\rm d}\epsilon_{a}}{{\rm d}t}
% + \frac{\rhog}{\rho_{a}} \frac{{\rm d}u_{a}}{{\rm d}t}
%\Bigg\} = 0.
%\label{eq:dedt}
%\end{align}

Given that we have already determined the form of the respective terms in the acceleration and dust-to-gas ratio evolution equations, we can use these terms to constrain the corresponding terms in the ${\rm d}\deltav/{\rm d}t$ and ${\rm d}u/{\rm d}t$ equations.

\subsubsection{Drag terms}
The drag term in Eq.~\ref{eq:genmomentum_deltav} causes differential velocity of the fluids to be dissipated into heat. From Eq.~\ref{eq:dedt} we have
\begin{equation}
\sum_{a} m_{a} \left(1 - \epsda \right)\left( \frac{{\rm d}u_{a}}{{\rm d}t}\right)_{\rm drag} = - \sum_{a} m_{a}\epsda \left(1 - \epsda \right) \deltav_{a}\cdot \left( \frac{{\rm d}{\deltav}_{a}}{{\rm d}t} \right)_{\rm drag},
\end{equation}
%WITH DTG
%\begin{equation}
%\sum_{a} m_{a} \frac{\rhog}{\rho_{a}} \left( \frac{{\rm d}u_{a}}{{\rm d}t}\right)_{\rm drag} = - \sum_{a} m_{a} \frac{\rhog\rhod}{\rho_{a}^{2}} \deltav_{a}\cdot \left( \frac{{\rm d}{\deltav}_{a}}{{\rm d}t} \right)_{\rm drag},
%\end{equation}
giving
\begin{equation}
\sum_{a} m_{a} \left(1 - \epsda \right)\left( \frac{{\rm d}u_{a}}{{\rm d}t}\right)_{\rm drag} =\sum_{a} m_{a} \epsda \left(1 - \epsda \right)  \frac{\deltav_{a}^{2}}{t_{\rm s}},
\end{equation}
%WITH DTG
%\begin{equation}
%\sum_{a} m_{a} \frac{\rhog}{\rho_{a}} \left( \frac{{\rm d}u_{a}}{{\rm d}t}\right)_{\rm drag} = \sum_{a} m_{a} \frac{\rhog\rhod}{\rho_{a}^{2}} \frac{\deltav_{a}^{2}}{t_{\rm s}},
%\end{equation}
and thus, as expected,
\begin{equation}
\left( \frac{{\rm d}u_{a}}{{\rm d}t}\right)_{\rm drag} = \epsda \frac{\deltav_{a}^{2}}{t_{\rm s}},
\end{equation}
%WITH DTG
%\begin{equation}
%\left( \frac{{\rm d}u_{a}}{{\rm d}t}\right)_{\rm drag} = \frac{\rhod}{\rho_{a}} \frac{\deltav_{a}^{2}}{t_{\rm s}},
%\end{equation}
consistent with Eq.~\ref{eq:newusingle}.

\subsubsection{${\bf f}_{\rm gas}$ and $P{\rm d}$V work terms}
 For the usual case of hydrodynamics, the pressure force on the gas gives rise to a $P{\rm d}V$ work term in the energy equation (Eq.~\ref{eq:newusingle}). In the case of dusty gas the same situation arises, but by a combination of the force terms in the ${\rm d}{\bf v}/{\rm d}t$ and ${\rm d}\deltav/{\rm d}t$ equations balancing the term in the energy equation. Substituting the force terms from Eqs.~\ref{eq:genmomentum_bary} and \ref{eq:genmomentum_deltav} in Eq.~\ref{eq:dedt}, we have
\begin{align}
\sum_{a} m_{a} \left(1 - \epsda \right) \left( \frac{{\rm d}u_{a}}{{\rm d}t}\right)_{P{\rm d}V} =  - \sum_{a} m_{a} & \left[\left(1 - \epsda \right){\bf v}_{a}\cdot {\bf f}_{\rm g} \right. \nonumber \\
& \left.- \epsda \left(1 -\epsda \right) \deltav_{a}\cdot {\bf f}_{\rm g} \right], \nonumber \\
=  - \sum_{a} m_{a} &\left(1 - \epsda \right){\bf v}_{{\rm g},a}\cdot {\bf f}_{\rm g}.
\label{eq:pdvterm}
\end{align}

 %WITH DTG
%\begin{align}
%\sum_{a} m_{a} \frac{\rhog}{\rho_{a}} \left( \frac{{\rm d}u_{a}}{{\rm d}t}\right)_{P{\rm d}V} & =  - \sum_{a} m_{a} \left[\frac{\rhog}{\rho_{a}}{\bf v}_{a}\cdot {\bf f}_{\rm g} - \frac{\rhog\rhod}{\rho_{a}^{2}} \deltav_{a}\cdot {\bf f}_{\rm g} \right], \nonumber \\
%& =  - \sum_{a} m_{a} \frac{\rhog}{\rho_{a}}{\bf v}_{{\rm g},a}\cdot {\bf f}_{\rm g}.
%\label{eq:pdvterm}
%\end{align}
Using the expression for ${\bf f}_{\rm g}$ implied by Eq.~\ref{eq:sphmom} we have
\begin{align}
\sum_{a} m_{a} \left(1 - \epsda \right)\left( \frac{{\rm d}u_{a}}{{\rm d}t}\right)_{P{\rm d}V} & = \sum_{a} m_{a} \sum_{b} m_{b} \frac{P_{a}}{\Omega_{a} \rho_{a}^{2}} {\bf v}_{{\rm g},a}. \nabla W_{ab}(h_{a}) \nonumber \\
&+ \sum_{a} m_{a} \sum_{b} m_{b} \frac{P_{b}}{\Omega_{b} \rho_{b}^{2}}  {\bf v}_{{\rm g},a}.\nabla W_{ab}(h_{b}).
\end{align}

%WITH DTG
%\begin{align}
%\sum_{a} m_{a} \frac{\rhog}{\rho_{a}} \left( \frac{{\rm d}u_{a}}{{\rm d}t}\right)_{P{\rm d}V} & = \sum_{a} m_{a} \sum_{b} m_{b} \frac{P_{a}}{\Omega_{a} \rho_{a}^{2}} {\bf v}_{{\rm g},a}. \nabla W_{ab}(h_{a}) \nonumber \\
%&+ \sum_{a} m_{a} \sum_{b} m_{b} \frac{P_{b}}{\Omega_{b} \rho_{b}^{2}}  {\bf v}_{{\rm g},a}.\nabla W_{ab}(h_{b}).
%\end{align}
Swapping summation indices in the second term, using the antisymmetry of the kernel gradient $\nabla W_{ba} = -\nabla W_{ab}$ and rearranging, we thus find that the corresponding term in the internal energy equation is given by
\begin{equation}
\left(\frac{{\rm d}u_{a}}{{\rm d}t}\right)_{P{\rm d}V} = \frac{P_{a}}{\Omega_{a} \rho_{a} \rho_{{\rm g},a}} \sum_{b} m_{b} \left({\bf v}_{{\rm g},a} - {\bf v}_{{\rm g},b} \right) \cdot \nabla W_{ab}(h_{a}),
\end{equation}
which is a generalisation of the usual $P{\rm d}V$ work term for hydrodynamics. This gives the numerical representation of the first term on the right hand side of Eq.~\ref{eq:newusingle}. Furthermore, it shows that the combination of this term with the pressure gradient terms in the acceleration and ${\rm d}\deltav/{\rm d}t$ equations conserves energy exactly.

\subsubsection{$-(\Delta{\bf v}.\nabla){\bf v}$ term}
 The SPH expression for the $-(\deltav.\nabla) {\bf v}$ term in Eq.~\ref{eq:genmomentum_deltav} is related to the numerical form of the anisotropic pressure term in Eq.~\ref{eq:genmomentum_bary}. This is entirely analogous to the situation in smoothed particle magnetohydrodynamics, in which the SPH expression for the $({\bf B}.\nabla) {\bf v}$ term in the induction equation for the magnetic field constrains the form of the anisotropic stress in the momentum equation (see \citealt{pricemonaghan04b}; in that case with opposite sign to ours, resulting in tension perpendicular to magnetic field lines).

Starting with the momentum-conserving expression for the anisotropic pressure term, we have a balance between these two terms of the form
\begin{align}
\sum_{a}m_{a} \epsda \left(1 - \epsda \right) & \deltav_{a}\cdot \frac{{\rm d}{\deltav}_{a}}{{\rm d}t} = -\sum_{a} m_{a} {\bf v}_{a}. \frac{{\rm d}{\bf v}_{a}}{{\rm d}t}, \nonumber \\
= & \sum_{a} m_{a}  \sum_{b} m_{b} \frac{\epsda \left( 1 - \epsda \right) {\bf v}_{a} \cdot\deltav_{a}}{\Omega_{a} \rho_{a}} \deltav_{a}.\nabla W_{ab}(h_{a}) \nonumber \\
+ & \sum_{a} m_{a} \sum_{b} m_{b} \frac{\epsdb \left( 1 - \epsdb \right){\bf v}_{a}\cdot \deltav_{b}}{\Omega_{b} \rho_{b}} \deltav_{b}.\nabla W_{ab}(h_{b}).
\end{align}
%WITH DTG
%\begin{align}
%\sum_{a} \frac{\rhog\rhod}{\rho_{a}^{2}} & \deltav_{a}\cdot \frac{{\rm d}{\deltav}_{a}}{{\rm d}t} = -\sum_{a} m_{a} {\bf v}_{a}. \frac{{\rm d}{\bf v}_{a}}{{\rm d}t}, \nonumber \\
%= & \sum_{a} m_{a}  \sum_{b} m_{b} \frac{\rhog\rhod {\bf v}_{a} \cdot\deltav_{a}}{\Omega_{a} \rho_{a}^{3}} \deltav_{a}.\nabla W_{ab}(h_{a}) \nonumber \\
%& +\sum_{a} m_{a} \sum_{b} m_{b} \frac{\rhog\rhod {\bf v}_{a}\cdot \deltav_{b}}{\Omega_{b} \rho_{b}^{3}} \deltav_{b}.\nabla W_{ab}(h_{b}).
%\end{align}
Swapping summation indices in the second term, using the antisymmetry of the kernel gradient and rearranging, we find that the corresponding term in ${\rm d}\deltav/{\rm d}t$ is given by
\begin{equation}
-(\deltav.\nabla){\bf v} = \frac{1}{\rho_{a}\Omega_{a}}\sum_{b} m_{b} ({\bf v}_{a} - {\bf v}_{b}) \deltav_{a}.\nabla W_{ab} (h_{a}).
\end{equation}
As with the $P{\rm d}V$ terms above, this is merely a demonstration of the conjugate nature of the differencing and symmetric SPH derivative operators, as discussed by a number of authors, including \citet{cumminsrudman99,price10,price12} and \citet{triccoprice12}.

\subsubsection{$(\Delta{\bf v}.\nabla)u$ and $\nabla ((\rhod-\rhog) \Delta{\bf v}^{2}/\rho)$ terms}
The remaining terms in the internal energy and ${\rm d}\deltav/{\rm d}t$ equations are constrained by the requirement of energy conservation (Eq.~\ref{eq:dedt}) via the balance with the SPH form of Eq.~\ref{eq:gendtgevol}. The last two terms in Eq.~\ref{eq:dedt} provide a constraint of the form
\begin{equation}
\sum_{a} m_{a}\left(1 - \epsda \right) \left( \frac{{\rm d}u_{a}}{{\rm d}t}\right)_{\rm term} = \sum_{a} m_{a} u_{a} \frac{{\rm d}\epsilon_{a}}{{\rm d}t}.
\end{equation}
%WITH DTG
%\begin{equation}
%\sum_{a} m_{a} \frac{\rhog}{\rho_{a}} \left( \frac{{\rm d}u_{a}}{{\rm d}t}\right)_{\rm term} = \sum_{a} m_{a} u_{a} \frac{\rhog^{2}}{\rho_{a}^{2}}\frac{{\rm d}\epsilon_{a}}{{\rm d}t}.
%\end{equation}
Substitution of Eq.~\ref{eq:dusttogassph} and rearrangement of the double summation as performed previously, we find a corresponding term in the energy equation in the form
\begin{equation}
\left( \frac{{\rm d}u_{a}}{{\rm d}t}\right)_{\rm term} = - \frac{\epsda}{\Omega_{a} \rho_{a}}  \sum_{b} m_{b} (u_{a} - u_{b}) \deltav_{a} .\nabla W_{ab} (h_{a}).
\end{equation}
%WITH DTG
%\begin{equation}
%\left( \frac{{\rm d}u_{a}}{{\rm d}t}\right)_{\rm term} = - \frac{\rhod^{a}}{\Omega_{a} \rho_{a}^{2}}  \sum_{b} m_{b} (u_{a} - u_{b}) \deltav_{a} .\nabla W_{ab} (h_{a}).
%\end{equation}
This is indeed an SPH representation of $\rhod/\rho (\deltav.\nabla) u$, as expected from Eq.~\ref{eq:newusingle}.
%\begin{align}
%\sum_{a} m_{a} \frac{\rhog}{\rho_{a}} \left( \frac{{\rm d}u_{a}}{{\rm d}t}\right)_{\rm term} 
%&= -\sum_{a} m_{a} u_{a} \frac{1}{\rho_{{\rm g},a}} \sum_{b} m_{b} \frac{\rhod^{a} \deltav_{a}}{\Omega_{a} \rho_{a}} .\nabla W_{ab} (h_{a}) \nonumber \\
%& - \sum_{a} m_{a} u_{a} \frac{\rho_{a}^{2}}{\rho_{{\rm g},a}^{2}} \sum_{b} m_{b} \frac{\rhog^{b}\rhod^{b} \deltav_{b}}{\Omega_{b} \rho_{b}^{3}}\cdot \nabla W_{ab} (h_{b}).  \nonumber \\
%&= -\sum_{a} m_{a} u_{a} \frac{\rhod^{a}}{\rho_{{\rm g},a}} \sum_{b} m_{b} \frac{\deltav_{a}}{\Omega_{a} \rho_{a}} .\nabla W_{ab} (h_{a}) \nonumber \\
%& + \sum_{b} m_{b} u_{b} \frac{\rho_{b}^{2}}{\rho_{{\rm g},b}^{2}} \sum_{a} m_{a} \frac{\rhog^{a}\rhod^{a} \deltav_{a}}{\Omega_{a} \rho_{a}^{3}}\cdot \nabla W_{ab} (h_{a}). 
%\end{align}

The remaining term in the ${\rm d}\deltav/{\rm d}t$ equation is constrained in a similar way. From Eq.~\ref{eq:dedt} we have
\begin{equation}
\sum_{a} m_{a} \epsda \left(1 - \epsda \right) \deltav_{a}\cdot \left(\frac{{\rm d}{\deltav}_{a}}{{\rm d}t} \right)_{\rm term} = -\sum_{a} m_{a} \left(1 - 2 \epsda \right) \frac12 \deltav_{a}^{2} \frac{{\rm d}\epsilon_{a}}{{\rm d}t},
\end{equation}
%WITH DTG
%\begin{equation}
%\sum_{a} m_{a} \frac{\rhog\rhod}{\rho_{a}^{2}} \deltav_{a}\cdot \left(\frac{{\rm d}{\deltav}_{a}}{{\rm d}t} \right)_{\rm term} = -\sum_{a} m_{a} \frac{\rhog^{2}}{\rho_{a}^{2}} \frac{1 - \epsilon_{a}}{1 + \epsilon_{a}} \frac12 \deltav_{a}^{2} \frac{{\rm d}\epsilon_{a}}{{\rm d}t},
%\end{equation}
giving, by substitution of Eq.~\ref{eq:dusttogassph} and rearrangement of the double summation, a corresponding term in ${\rm d}\deltav/{\rm d}t$ of the form
\begin{equation}
\left(\frac{{\rm d}{\deltav}_{a}}{{\rm d}t} \right)_{\rm term} = \frac{1}{2\rho_{a}\Omega_{a}} \sum_{b} m_{b} \left[ \left(1 - 2 \epsda \right) \deltav_{a}^{2} -  \left(1 - 2 \epsdb \right) \deltav_{b}^{2} \right] \nabla W_{ab} (h_{a}).
\end{equation}
%WITH DTG
%\begin{equation}
%\left(\frac{{\rm d}{\deltav}_{a}}{{\rm d}t} \right)_{\rm term} = \frac{1}{2\rho_{a}\Omega_{a}} \sum_{b} m_{b} \left[ \frac{1 - \epsilon_{a}}{1 + \epsilon_{a}} \deltav_{a}^{2} -  \frac{1 - \epsilon_{b}}{1 + \epsilon_{b}} \deltav_{b}^{2} \right] \nabla W_{ab} (h_{a}).
%\end{equation}
This completes the formulation of all of the non-dissipative part of the algorithm, as well as the physical dissipation due to drag. The only major issue remaining is to generalise the usual SPH artificial viscosity and conductivity terms in an appropriate manner for the mixture. 

%Choice for the $\deltav^{2}$ term to conserve exactly the total kinetic energy of the two fluids?
%Comparison with the two-fluid approach: no need for interpolating delta v over the neighbours (local value of deltav, easier for implicit, get rid of the drag kernels)

\subsection{Artificial viscosity and conductivity terms}
\label{sec:diss}
 Since viscosity and conductivity terms apply only to the gas we require that a) they involve only gas properties, e.g. the gas velocity rather than the barycentric velocity; b) conserve momentum; and c) conserve energy.
 
\subsubsection{Artificial viscosity}
From Eq.~\ref{eq:genmomentum_bary} we have
\begin{equation}
\left(\frac{{\rm d} \vb_{a}}{{\rm d} t}\right)_{\rm visc} =  \frac{\rhog^{a}}{\rho_{a}}{\bf f}_{\rm g, visc}.
\end{equation}
A naive approach would be to simply insert the usual \citep{monaghan97,price12} expression for the artificial viscosity term as ${\rm f}_{\rm g, visc}$, giving
\begin{equation}
\left(\frac{{\rm d} \vb_{a}}{{\rm d} t}\right)_{\rm visc} =  \frac{\rhog^{a}}{\rho_{a}} \sum_{b} m_{b} \frac{v_{\rm sig}}{\overline{\rho}_{ab}} ({\bf v}^{\rm g}_{a} - {\bf v}^{\rm g}_{b}) \cdot \hat{\bf r}_{ab}\overline{\nabla W}_{ab}.
\label{eq:avnaive}
\end{equation}
where $\overline{\nabla W}_{ab} \equiv \frac12 [\nabla W_{ab} (h_{a}) + \nabla W_{ab} (h_{b})]$, $\overline{\rho}_{ab} = \frac12(\rho_{a} + \rho_{b})$ and $v_{\rm sig}$ is the maximum speed of signal propagation. For the case of hydrodynamics the pairwise signal speed is given by
\begin{equation}
v_{\rm sig} = \alpha \frac{( c_{{\rm s},a} + c_{{\rm s},b})}{2} + \beta\vert {\bf v}^{\rm g}_{a} - {\bf v}_{b}^{\rm g}\vert,
\end{equation}
where $\alpha$ and $\beta$ are the usual linear and quadratic SPH viscosity parameters, and the term is only applied where $({\bf v}^{\rm g}_{a} - {\bf v}^{\rm g}_{b})\cdot\hat{\bf r}_{ab} < 0$.

The problem with Eq.~\ref{eq:avnaive} is that it does not conserve momentum, which requires (c.f. Eq.~\ref{eq:momcons}) that the term in the ${\rm d}{\bf v}/{\rm d}t$ equation is antisymmetric with respect to the particle labels $a$ and $b$. The simplest approach is to symmetrise the $\rhog/\rho$ term, giving
\begin{equation}
\left(\frac{{\rm d} \vb_{a}}{{\rm d} t}\right)_{\rm visc} = \sum_{b} m_{b} \frac12 \left(\frac{\rhog^{a}}{\rho_{a}}  + \frac{\rhog^{b}}{\rho_{b}}\right)\frac{v_{\rm sig}}{\overline{\rho}_{ab}} ({\bf v}^{\rm g}_{a} - {\bf v}^{\rm g}_{b}) \cdot \hat{\bf r}_{ab}\overline{\nabla W}_{ab},
\label{eq:avgood}
\end{equation}
and thus implying
\begin{equation}
{\bf f}_{\rm g, visc} = \frac{\rho_{a}}{\rhog^{a}} \sum_{b} m_{b} \frac12 \left(\frac{\rhog^{a}}{\rho_{a}}  + \frac{\rhog^{b}}{\rho_{b}}\right)\frac{v_{\rm sig}}{\overline{\rho}_{ab}} ({\bf v}^{\rm g}_{a} - {\bf v}^{\rm g}_{b}) \cdot \hat{\bf r}_{ab}\overline{\nabla W}_{ab},
\end{equation}
or equivalently
\begin{equation}
{\bf f}_{\rm g, visc} = \frac{\rho_{a}}{\rhog^{a}} \sum_{b} m_{b} \left(1 - \overline{\epsd}_{ab}\right)\frac{v_{\rm sig}}{\overline{\rho}_{ab}} ({\bf v}^{\rm g}_{a} - {\bf v}^{\rm g}_{b}) \cdot \hat{\bf r}_{ab}\overline{\nabla W}_{ab},
\end{equation}
where $\overline{\epsd}_{ab} = \frac12 (\epsda + \epsd_{b})$. The corresponding term in the energy equation is constrained, from Eq.~\ref{eq:pdvterm}, according to
\begin{align}
\sum_{a} m_{a} \frac{\rhog}{\rho_{a}} \left( \frac{{\rm d}u_{a}}{{\rm d}t}\right)_{\rm visc}
& =  - \sum_{a} m_{a} \frac{\rhog}{\rho_{a}}{\bf v}_{{\rm g},a}\cdot {\bf f}_{\rm g, visc},
\label{eq:pdvtermav}
\end{align}
giving
\begin{equation}
\left( \frac{{\rm d}u_{a}}{{\rm d}t}\right)_{\rm visc} = -\frac{\rho_{a}}{\rhog^{a}} \sum_{b} m_{b} \frac{v_{\rm sig}}{\overline{\rho}_{ab}} \frac12 ({\bf v}^{\rm g}_{ab}\cdot\hat{\bf r}_{ab})^{2} \overline{F}_{ab},
\label{eq:dudtvisc}
\end{equation}
where $F_{ab}$ is defined such that $\nabla W_{ab} = \hat{\bf r}_{ab} F_{ab}$. The viscous heating term is also only applied where $({\bf v}^{\rm g}_{a} - {\bf v}^{\rm g}_{b})\cdot\hat{\bf r}_{ab} < 0$.

\subsubsection{Artificial conductivity}
The artificial conductivity term, required to treat discontinuities in the thermal energy \citep{price08} must satisfy the constraint of energy conservation, i.e.
\begin{equation}
\sum_{a} m_{a} \frac{\rhog^{a}}{\rho_{a}} \left( \frac{{\rm d}u_{a}}{{\rm d}t}\right)_{\rm cond} = 0.
\end{equation}
A simple generalisation of the usual artificial conductivity formulation that satisfies this constraint is given by
\begin{equation}
\left( \frac{{\rm d}u_{a}}{{\rm d}t}\right)_{\rm cond} =  \frac{\rho_{a}}{\rhog^{a}}\sum_{b} m_{b} \frac{\alpha_{u} v_{\rm sig, u}}{\overline{\rho}_{ab}} (u_{a} - u_{b}) \overline{F}_{ab},
\label{eq:dudtcond}
\end{equation}
where $\alpha_{u}\sim 1$ is the dimensionless artificial conductivity parameter and $v_{\rm sig, u}$ is the signal speed used for the conductivity. The standard choices for $v_{\rm sig, u}$ are either $v_{\rm sig, u} = \sqrt{\vert P_{a} - P_{b}\vert / \overline{\rho}_{ab}}$ proposed by \citet{price08} or $v_{\rm sig, u} = \vert {\bf v}_{ab} \cdot \hat{\bf r}_{ab} \vert$ proposed by \citet{wadsleyetal08}. We use the former by default, adopting the latter in simulations involving gravity. The proof that the standard artificial conductivity term results in a positive definite contribution to the entropy can be found in \citet{pricemonaghan04a}.

\subsubsection{Artificial dissipation in $\deltav$}
\label{sec:dissdeltav}
The dissipation terms given above are based on physical considerations. One may alternatively consider that artificial dissipation is introduced only to treat jumps and discontinuities (for example in the case of an inviscid gas). Hence, alternative formulations for the dissipation term may be possible, where the only requirements are that the scheme should conserve momentum and (total) energy and result in a positive definite contribution to the entropy. Indeed, with the physical formulation above we found that some post-shock oscillations remained in the weak-coupling limit (i.e. large $\deltav$).

 We found that these oscillations could be eliminated by adding an additional dissipative term to the $\deltav$ equation of the form
\begin{equation}
\left(\frac{{\rm d}\deltav}{{\rm d}t}\right)_{\rm diss} = \frac{1}{\epsd_{a}(1 - \epsd_{a})}\sum_{b} m_{b} \frac{v_{\rm sig, \deltav}}{\overline{\rho}_{ab}} \overline{\epsd}_{ab}\left(1 - \overline{\epsd}_{ab}\right)(\deltav_{a}- \deltav_{b})\cdot \hat{\bf r}_{ab} \overline{\nabla W}_{ab},
\label{eq:deltavdiss}
\end{equation}
where $v_{\rm sig, \deltav}$ is a signal speed that may or may not equal the one used in the viscosity term, with a corresponding heating term of the form
\begin{equation}
\left(\frac{{\rm d}u}{{\rm d}t}\right)_{\rm diss} = -\frac{\rho_{a}}{\rhog^{a}}\sum_{b} m_{b} \frac{v_{\rm sig, \deltav}}{\overline{\rho}_{ab}} \overline{\epsd}_{ab}\left(1 - \overline{\epsd}_{ab}\right) \frac12\left[(\deltav_{a} - \deltav_{b})\cdot\hat{\bf r}_{ab}\right]^{2} \overline{F}_{ab},
\end{equation}
which is positive-definite since $\overline{F}_{ab}$ is negative-definite for standard kernels. This term is effectively a non-linear drag, which converts gradients in $\deltav$, but not $\deltav$ itself, into heat. We found that an effective and appropriate choice for $v_{\rm sig, \deltav}$ was given by
\begin{equation}
v_{\rm sig, \deltav} = \alpha_{\deltav} \vert \deltav_{a} - \deltav_{b})\cdot\hat{\bf r}_{ab} \vert,
\end{equation}
where $\alpha_{\deltav}$ is a dimensionless coefficient of order unity. This produces a second-order dissipation in $\deltav$ analogous to the $\beta$ term in the usual SPH viscosity. We also found that the dissipation associated with this term could be reduced further by only applying this term only when $(\deltav_{a} - \deltav_{b})\cdot\hat{\bf r}_{ab} < 0$.
 
\subsection{Alternative dissipation formulation}
\label{sec:diss-nonconserv}
 We also investigated a number of alternative dissipation schemes. In every case the dissipation reduced to the usual shock-capturing terms in SPH in the gas-only limit, so differences between schemes only became apparent when $\deltav$ was large, i.e. in the weak-coupling limit. By experimenting with a number of possible formulations (including several derived in a manner analogous to \citealt{monaghan97} and \citealt{pricemonaghan04a,pricemonaghan05}), we obtained our best results on the zero-drag \textsc{dustyshock} problem with dissipation terms of the form
\begin{eqnarray}
\left(\frac{{\rm d}{\bf v}}{{\rm d}t}\right)_{\rm diss} & = & \sum_{b} m_{b}\frac{v_{\rm sig}}{\overline{\rho}_{ab}} ({\bf v}^{\rm g}_{a} - {\bf v}^{\rm g}_{b}) \cdot \hat{\bf r}_{ab}\overline{\nabla W}_{ab}, \label{eq:altdiss1} \\
\left(\frac{{\rm d}\deltav}{{\rm d}t}\right)_{\rm diss} & = & \frac{\rho_{a}}{\rhog^{a}}\sum_{b} m_{b} \frac{v_{\rm sig}}{\overline{\rho}_{ab}} (\deltav_{a}- \deltav_{b})\cdot \hat{\bf r}_{ab} \overline{\nabla W}_{ab}.
\label{eq:altdiss2}
\end{eqnarray}
where the conductivity and viscous heating terms are as previously derived (Eqs.~\ref{eq:dudtcond} and \ref{eq:dudtvisc}, respectively). The main difference to the standard formulation is that the term in the ${{\rm d}{\bf v}}/{{\rm d}t}$ equation is not included in the $\deltav$ equation as part of ${\bf f}_{\rm gas}$. This means that the formulation, retaining Eq.~\ref{eq:dudtvisc} as the heating term, is non-conservative (the alternative is to use a matching heating term but which does not result in a positive definite entropy contribution --- we also tried this and it does not change the results). Our various attempts to produce a conservative formulation based on the above dissipation terms all produced worse results, which can be seen by comparing the results from this formulation with our ``standard'' formulation of dissipative terms (compare Figs.~\ref{fig:dustyshock-weak} and \ref{fig:dustyshock-alt}). However, the error in energy conservation using Eqs.~\ref{eq:altdiss1}--\ref{eq:altdiss2}, whilst not being rigourously equal to zero, was found to be small ($\sim 10^{-6}$) for the test problems studied in this paper.

%  For example, we could instead formulate the dissipation terms in the manner described by \citet{monaghan97} (for hydrodynamics) or by \citet{pricemonaghan04a,pricemonaghan05} (for magnetohydrodynamics). Here the approach is to construct a dissipation term in the specific energy equation 

\subsection{Summary}
Summarising, the SPH representation of Eqs.~\ref{eq:genmass_rho}--\ref{eq:newusingle}, with all terms including the conservative shock-capturing terms, are given by
\begin{align}
\rho_{a} & = \sum_{b} m_{b} W_{ab} (h_{a}), \label{eq:density-final} \\
\frac{{\rm d}\epsda}{{\rm d} t} & = - \sum_{b} m_{b} \left[ \frac{\epsda \left(1 - \epsda \right)}{\Omega_{a} \rho_{a}}  \deltav_{a}\cdot\nabla_{a} W_{ab} (h_{a})\right. \nonumber \\
 & \hspace{1.2cm} + \left. \frac{\epsdb \left(1 - \epsdb \right)}{\Omega_{b} \rho_{b}}\deltav_{b}\cdot \nabla_{a} W_{ab} (h_{b}) \right], \label{eq:dusttogas-final} \\
\frac{{\rm d} \vb_{a}}{{\rm d} t} = & \phantom{+}(1 - \epsda){\bf f}_{\rm g}   + {\bf f}_{a} \nonumber\\
& -\sum_{b} m_{b} \left[ \frac{\epsda\left(1 - \epsda \right) \deltav_{a}}{\Omega_{a} \rho_{a}} \deltav_{a}.\nabla W_{ab}(h_{a}) \right. \nonumber \\
&  \phantom{\sum_{b} m_{b}[} + \left.\frac{\epsdb\left(1 - \epsdb \right)\deltav_{b}}{\Omega_{b} \rho_{b}} \deltav_{b}.\nabla W_{ab}(h_{b}) \right] 
\label{eq:mom-final} \\
\frac{{\rm d}{\deltav}_{a}}{{\rm d}t} & =  -\frac{\deltav_{a}}{t_{\rm s, a}}  - {\bf f}_{\rm g} \nonumber \\
& +\frac{1}{\rho_{a}\Omega_{a}}\sum_{b} m_{b} ({\bf v}_{a} - {\bf v}_{b}) \deltav_{a}.\nabla W_{ab} (h_{a}) \label{eq:deltav-final} \\
& +\frac{1}{2\rho_{a}\Omega_{a}} \sum_{b} m_{b} \left[ \left(1 - 2 \epsda \right) \deltav_{a}^{2} -  \left(1 - 2 \epsdb \right) \deltav_{b}^{2} \right] \nabla W_{ab} (h_{a}), \nonumber \\ 
& + \frac{1}{\epsd_{a}(1 - \epsd_{a})}\sum_{b} m_{b} \frac{v_{\rm sig, \deltav}}{\overline{\rho}_{ab}} \overline{\epsd}_{ab}\left(1 - \overline{\epsd}_{ab}\right)(\deltav_{a}- \deltav_{b})\cdot \hat{\bf r}_{ab} \overline{\nabla W}_{ab},\nonumber 
\end{align}
where
\begin{align}
(1 - \epsda){\bf f}_{\rm g} & = -\sum_{b} m_{b} \left[ \frac{P_{a}}{\Omega_{a} \rho_{a}^{2}} \nabla W_{ab}(h_{a}) + \frac{P_{b}}{\Omega_{b} \rho_{b}^{2}} \nabla W_{ab}(h_{b}) \right], \nonumber \\
& + \sum_{b} m_{b} \left(1 - \overline{\epsd}_{ab}\right)\frac{v_{\rm sig}}{\overline{\rho}_{ab}} ({\bf v}^{\rm g}_{a} - {\bf v}^{\rm g}_{b}) \cdot \hat{\bf r}_{ab}\overline{\nabla W}_{ab},
\end{align}
along with
\begin{align}
\frac{{\rm d}{\bf x}_{a}}{{\rm d}t} & = {\bf v}_{a}, \label{eq:dxdt-final}\\
t_{\rm s, a} & = \frac{\epsda ( 1 - \epsda) \rho_{a} }{K_{a}},
\end{align}
and
\begin{align}
\frac{{\rm d}u_{a}}{{\rm d}t} & = \frac{P_{a}}{\Omega_{a} \rho_{a} \rho_{{\rm g},a}} \sum_{b} m_{b} \left({\bf v}_{{\rm g},a} - {\bf v}_{{\rm g},b} \right) \cdot \nabla W_{ab}(h_{a}) \nonumber \\
& - \frac{\epsda}{\Omega_{a} \rho_{a}}  \sum_{b} m_{b} (u_{a} - u_{b}) \deltav_{a} .\nabla W_{ab} (h_{a}) \nonumber \\
& -\frac{\rho_{a}}{\rhog^{a}} \sum_{b} \frac{m_{b}}{\overline{\rho}_{ab}} \left[ v_{\rm sig} \frac12 ({\bf v}^{\rm g}_{ab}\cdot\hat{\bf r}_{ab})^{2} - \alpha_{u} v_{\rm sig, u} (u_{a} - u_{b}) \right] \overline{F}_{ab} \nonumber \\
& -\frac{\rho_{a}}{\rhog^{a}}\sum_{b} \frac{m_{b}}{\overline{\rho}_{ab}} v_{\rm sig, \deltav} \overline{\epsd}_{ab}\left(1 - \overline{\epsd}_{ab}\right) \frac12\left[(\deltav_{a} - \deltav_{b})\cdot\hat{\bf r}_{ab}\right]^{2} \overline{F}_{ab}, \nonumber \\
& + \epsda \frac{\deltav_{a}^{2}}{t_{\rm s, a}}. \label{eq:dudt-final}
\end{align}

\subsection{Timestepping}
 Once the SPH formulation of the spatial derivative terms has been specified, the choice of timestepping boils down to one's preferred method for solving the ordinary differential equations for ${\bf x}$, ${\bf v}$, $\epsd$, $\deltav$ and $u$, namely Eqs.~\ref{eq:dusttogas-final}--\ref{eq:dxdt-final} and, depending on the equation of state, Eq.~\ref{eq:dudt-final}. Any standard scheme can be used.

\subsubsection{Explicit integration}
For the tests in this paper we use an adapted version of the leapfrog integrator in the Velocity-Verlet form, where the predictor step is given by
\begin{align}
{\bf x}^{n+1} & = {\bf x}^{n} + \Delta t {\bf v}^{n} + \frac{(\Delta t)^{2}}{2}  \left(\frac{{\rm d}{\bf v}}{{\rm d}t}\right)^{n}, \\
{\bf v}^{*} & = {\bf v}^{n} + \Delta t \left(\frac{{\rm d}{\bf v}}{{\rm d}t}\right)^{n}, \\
\epsd^{*} & = \epsd^{n} + \Delta t \left(\frac{{\rm d}\epsd}{{\rm d}t}\right)^{n}, \\
\deltav^{*} & = \deltav^{n} + \Delta t \left(\frac{{\rm d}\deltav}{{\rm d}t}\right)^{n}, \\
u^{*} & = u^{n} + \Delta t \left(\frac{{\rm d}u}{{\rm d}t}\right)^{n},
\end{align}
whereupon all derivatives are evaluated, and a trapezoidal corrector step is used:
\begin{align}
{\bf v}^{n+1} & = {\bf v}^{n} + \frac12 \Delta t \left[ \left(\frac{{\rm d}{\bf v}}{{\rm d}t}\right)^{n} + \left(\frac{{\rm d}{\bf v}}{{\rm d}t}\right)^{*}\right], \\
\epsd^{n+1} & = \epsd^{n} + \frac12 \Delta t \left[ \left(\frac{{\rm d}\epsd}{{\rm d}t}\right)^{n} +  \left(\frac{{\rm d}\epsd}{{\rm d}t}\right)^{*}\right], \\
\deltav^{n+1} & = \deltav^{n} + \frac12 \Delta t \left[ \left(\frac{{\rm d}\deltav}{{\rm d}t}\right)^{n} +  \left(\frac{{\rm d}\deltav}{{\rm d}t}\right)^{*}\right], \\
u^{n+1} & = u^{n} + \frac12 \Delta t \left[ \left(\frac{{\rm d}u}{{\rm d}t}\right)^{n} + \left(\frac{{\rm d}u}{{\rm d}t}\right)^{*} \right].
\end{align}
This integrator is simple to implement, but will not be strictly second order accurate unless the position-dependent terms in the acceleration are dominant, since only one evaluation of the derivatives on the right hand side is used. We have therefore compared our results with those using a standard second-order Runge-Kutta integrator, requiring two derivative evaluations. We find no significant differences for the tests presented in this paper.

\subsubsection{Implicit integration of strong drag regimes}
\label{sec:implicit}
In the limit where the drag is strong, the explicit integration described above will be subject to the requirement $\Delta t < t_{\rm s}$, in addition to the usual Courant condition. This implies a prohibitive computational cost in strong drag regimes, where the stopping time is much shorter than the timescales in the problem of interest. In the two-fluid case, handling this requires a complicated implicit integration scheme due to the need to interpolate between different types of particles \citep{LP12b}. By contrast, there is only one set of particles and only one term involving $\ts$ in the one-fluid algorithm --- in the $\deltav$ equation, with a corresponding heating term in ${\rm d}u/{\rm d}t$ --- so implicit integration can be implemented in a straightforward manner using operator splitting.

Integration of the drag term involves the differential equation
\begin{equation}
\frac{\mathrm{d}\deltav}{\mathrm{d}t} = -\frac{\deltav}{\ts} + \mathbf{a}_{0} ,
\label{eq:simple_drag}
\end{equation}
where $\ts$ may in general be a function of $\deltav$ (for non-linear drag regimes) and $\mathbf{a}_{0} = \left(\mathrm{d}\deltav/\mathrm{d}t\right)_{0}^{n}$ is a constant. Starting with linear drag regimes for simplicity, the exact solution of Eq.~\ref{eq:simple_drag} is
\begin{equation}
\deltav(t) = \deltav(t_{0})e^{-t/\ts} + \mathbf{a}_{0}\ts \left(1 - e^{-t/\ts} \right) ,
\end{equation}
meaning that the evolution over a time step is given by
\begin{equation}
\deltav^{n + 1} = \deltav^{n}e^{-\Delta t/\ts} + \mathbf{a}_{0}\ts \left(1 - e^{-\Delta t/\ts} \right) .
\label{eq:timestep}
\end{equation}
In the case of weak drag regimes $(\Delta t \ll \ts)$, Eq.~\ref{eq:timestep} reduces to
\begin{equation}
\deltav^{n + 1} = - \deltav^{n} \frac{\Delta t}{\ts} + \mathbf{a}_{0}\Delta t + \mathcal{O}\left(\left(  \Delta t / \ts \right)^{2} \right),
\end{equation}
which is the expression used in the explicit time stepping above. For strong drag regimes $(\ts \ll \Delta t)$ Eq.~\ref{eq:timestep} gives
\begin{equation}
\deltav^{n+1}  =  \mathbf{a}_{0} \ts +\mathcal{O}\left( e^{-\frac{\Delta t}{\ts}}\right) ,
\label{eq;approx_strong}
\end{equation}
which is the terminal velocity approximation described in Paper~I.

 In practice, we can incorporate Eq.~\ref{eq:timestep} into the timestepping scheme described above as follows: First, compute $\mathbf{a}_{0}$ using all terms in $\mathrm{d}\deltav/\mathrm{d}t$ except for the drag. Second, compute $\deltav^{n + 1}_{\rm impl}$ according to Eq.~\ref{eq:timestep}. Finally, use this reconstruct the corresponding `explicit' derivative term, i.e.
\begin{equation}
\left(\frac{\mathrm{d}\deltav}{\mathrm{d}t}\right)_{\rm impl}^{n} =\frac{\left(\deltav^{n + 1}_{\rm impl} - \deltav^{n}  \right)}{\Delta t } ,
\end{equation}
This can then be used to specify the `explicit' drag term in a regular predictor-corrector scheme.

 The internal energy evolution is calculated according to
\begin{align}
u^{n+1} - u^{n} = & \frac{K}{\rhog} \int_{0}^{\Delta t} \left[ \deltav^{n} e^{-\frac{- t'}{\ts}} +  \mathbf{a}_{0}\ts \left(  1 -   e^{\frac{- t'}{\ts}}\right) \right]^{2}\mathrm{d}t' \nonumber \\
= & \frac{K \ts}{2 \rhog} \left\lbrace 2 \mathbf{a}_{0}^{2} \ts \Delta t  - e^{-2 \frac{\Delta t}{\ts}} \left( 1 -   e^{\frac{\Delta t}{\ts}} \right)  \right. \nonumber  \\
& \left. \times \left(\deltav^{n} - \mathbf{a}_{0}\ts \right)\left( \deltav^{n} - \mathbf{a}_{0}\ts + e^{\frac{\Delta t}{\ts}} \left(\deltav^{n} + 3 \mathbf{a}_{0}\ts  \right)  \right) \right\rbrace , \label{eq:uImpl_expr}
\end{align}
where $\dst K \ts /  \rhog = \epsilon$. The energy dissipated in the gas is positive since Eq.~\ref{eq:uImpl_expr} involves the integral of the positive function $\deltav (t)^{2}$. For $\Delta t \ll \ts$, Eq.~\ref{eq:uImpl_expr} reduces to the expected
\begin{equation}
u^{n+1} - u^{n} = \frac{K}{\rhog} \left( \deltav^{n}\right)^{2}\Delta t .
 \end{equation}
 For $\ts \ll \Delta t$, Eq.~\ref{eq:uImpl_expr} becomes 
 \begin{equation}
u^{n+1} - u^{n} =  \frac{K}{\rhog} \left(a_{0} \ts\right)^{2} \Delta t / \ts  ,
 \end{equation}
 implying that most of the energy is dissipated during the stationary state of the solution (the correction due to the transient regime is second order). A corresponding derivative term for use in the predictor-corrector can be calculated according to
\begin{equation}
\left(\frac{\mathrm{d}u}{\mathrm{d}t}\right)_{\rm impl}^{n} =\frac{\left(u^{n + 1} - u^{n}  \right)_{\rm impl}}{\Delta t } .
\label{eq:dudt_impl}
\end{equation}
Finally, the operator-split predictor step reads:
 \begin{align}
 \deltav^{**} & = \deltav^{n} + \Delta t \left(\frac{{\rm d}\deltav}{{\rm d}t}\right)_{0}^{n}, \\
 u^{**} & = u^{n} + \Delta t \left(\frac{{\rm d}u}{{\rm d}t}\right)_{0}^{n}, \\
 \deltav^{*} & = \deltav^{**} + \Delta t \left(\frac{\mathrm{d}\deltav}{\mathrm{d}t}\right)_{\rm impl}^{n}  , \\
 u^{*} & = u^{**} + \Delta t \left(\frac{\mathrm{d}u}{\mathrm{d}t}\right)_{\rm impl}^{n} ,
 \end{align}
 where the subscript $_{0}$ denotes all terms on the right hand side of Eqs.~\ref{eq:deltav-final} or \ref{eq:dudt-final} except those involving $\ts$, and we have used Eqs.~\ref{eq:timestep} and \ref{eq:dudt_impl} to compute the implicit drag terms. The corrector step is given by
 \begin{align}
\deltav^{**} & = \deltav^{n} + \frac12 \Delta t \left[ \left(\frac{{\rm d}\deltav}{{\rm d}t}\right)_{0}^{n} +  \left(\frac{{\rm d}\deltav}{{\rm d}t}\right)_{0}^{*}\right], \\
u^{**} & = u^{n} + \frac12 \Delta t \left[ \left(\frac{{\rm d}u}{{\rm d}t}\right)_{0}^{n} + \left(\frac{{\rm d}u}{{\rm d}t}\right)_{0}^{*} \right], \\
\deltav^{n+1} & = \deltav^{**} + \Delta t \left[  \left(\frac{\mathrm{d}\deltav}{\mathrm{d}t}\right)_{\rm impl}^{n} +  \left(\frac{\mathrm{d}\deltav}{\mathrm{d}t}\right)_{\rm impl}^{*}\right] , \\
u^{n+1} & = u^{**} + \Delta t \left[ \left(\frac{\mathrm{d}u}{\mathrm{d}t}\right)_{\rm impl}^{n} + \left(\frac{\mathrm{d}u}{\mathrm{d}t}\right)_{\rm impl}^{*} \right] .
\end{align}
The Lie truncation procedure used for the operator splitting above is consistent with a globally second order scheme. For non-linear drag regimes, the procedure is essentially the same as for linear drag, but may involve an additional numerical integration in the energy term. As an example, the calculations for the quadratic drag encountered at high Mach numbers in diluted gases or at high Reynolds numbers in a dense medium are given in Appendix~\ref{app:impquad}. The overall performance was found to be similar to the linear case.

%=======================================================================================================
\section{Tests}
\label{sec:tests}

We have implemented the one-fluid dust-and-gas SPH algorithm presented above in the public $N-$dimensional \textsc{ndspmhd} code \citep{price12}. A public version of the code that includes our implementation will be made available alongside this paper. Here, we benchmark the algorithm against the \textsc{dustybox}, \textsc{dustywave} and \textsc{dustyshock} problems described in \citet{LP11} as well as a new problem, \textsc{dustyoscill}, designed to probe the limitations of the one-fluid approach. The two-fluid counterparts of the first three tests were given in \citet{LP12a,LP12b}.

\subsection{\textsc{dustybox}}
\label{sec:dustybox}

\begin{figure}
\begin{center}
   \includegraphics[width=\columnwidth]{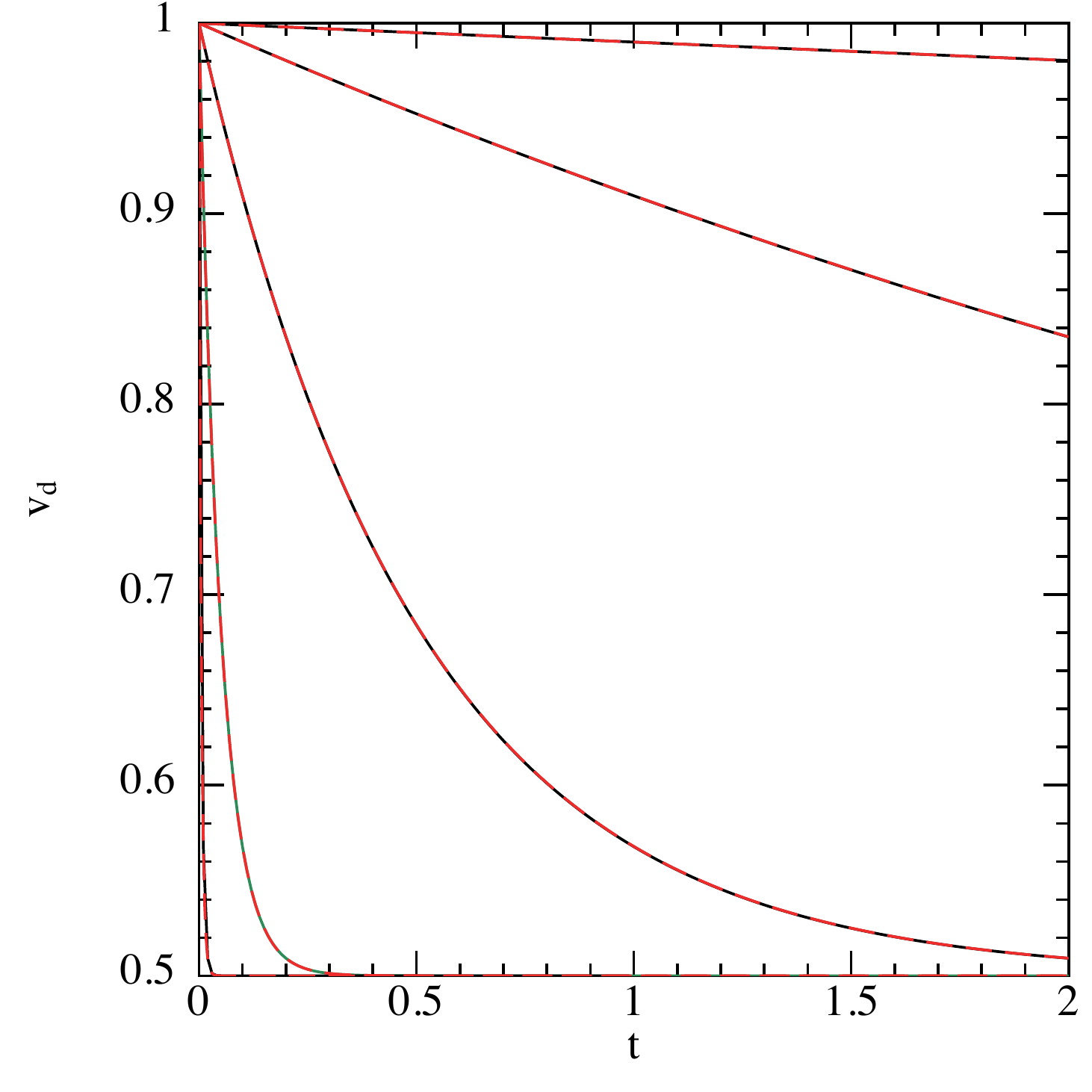}
   \caption{Dust velocity as a function of time in the \textsc{dustybox} problem with the one-fluid algorithm and implicit time integration, using $100$ particles and a dust-to-gas ratio of unity. A linear drag regime was used, with the drag coefficient given by $K = 0.01, 0.1, 1, 10$ and $100$ (top-to-bottom, solid/black lines). Results are in excellent agreement with the exact solution given by the long-dashed/red lines ($< 1\%$ error in $L_{1}$).}
\label{fig:K}
\end{center}
\end{figure}

The \textsc{dustybox} problem involves gas and dust moving with a constant differential velocity $\Delta v _{0} =  v _{d,0} -  v _{g,0}$ but where the barycentre of the mixture as a whole is at rest (i.e. momentum is only exchanged between the two phases via the drag term). The density as well as the dust fraction and the gas pressure are also constant. The test is mainly useful for checking the implementation of the drag terms, since the other terms are zero for this problem. With the one-fluid algorithm, the \textsc{dustybox} problem is essentially trivial. Indeed, when using the implicit scheme described in Sec.~\ref{sec:implicit}, we already use the exact solution in the integration procedure.

\subsubsection{Setup}

We setup the problem using $100$ particles in a 1D periodic domain --- half the number of particles compared to the two fluid algorithm. The gas sound speed was set to $c_{\rm s} =1$, the total density $\rho = 1$ and the dust fraction $\epsd = 0.5$ in code units, corresponding to an equal mixture of dust and gas. No artificial viscosity terms were applied. Although the two phases are drifting with respect to each other, the \textit{total} mass of the mixture remained constant as expected. In the \textsc{dustybox} problem, ${\bf v} = {\bf v}_{0} = 0$ (no advection of the total mass). During the simulation, we verified that the total linear and angular momentum as well as the total energy are exactly conserved, confirming that the implicit scheme works as expected.
 
\subsubsection{Results}
\begin{figure}
\begin{center}
   \includegraphics[width=\columnwidth]{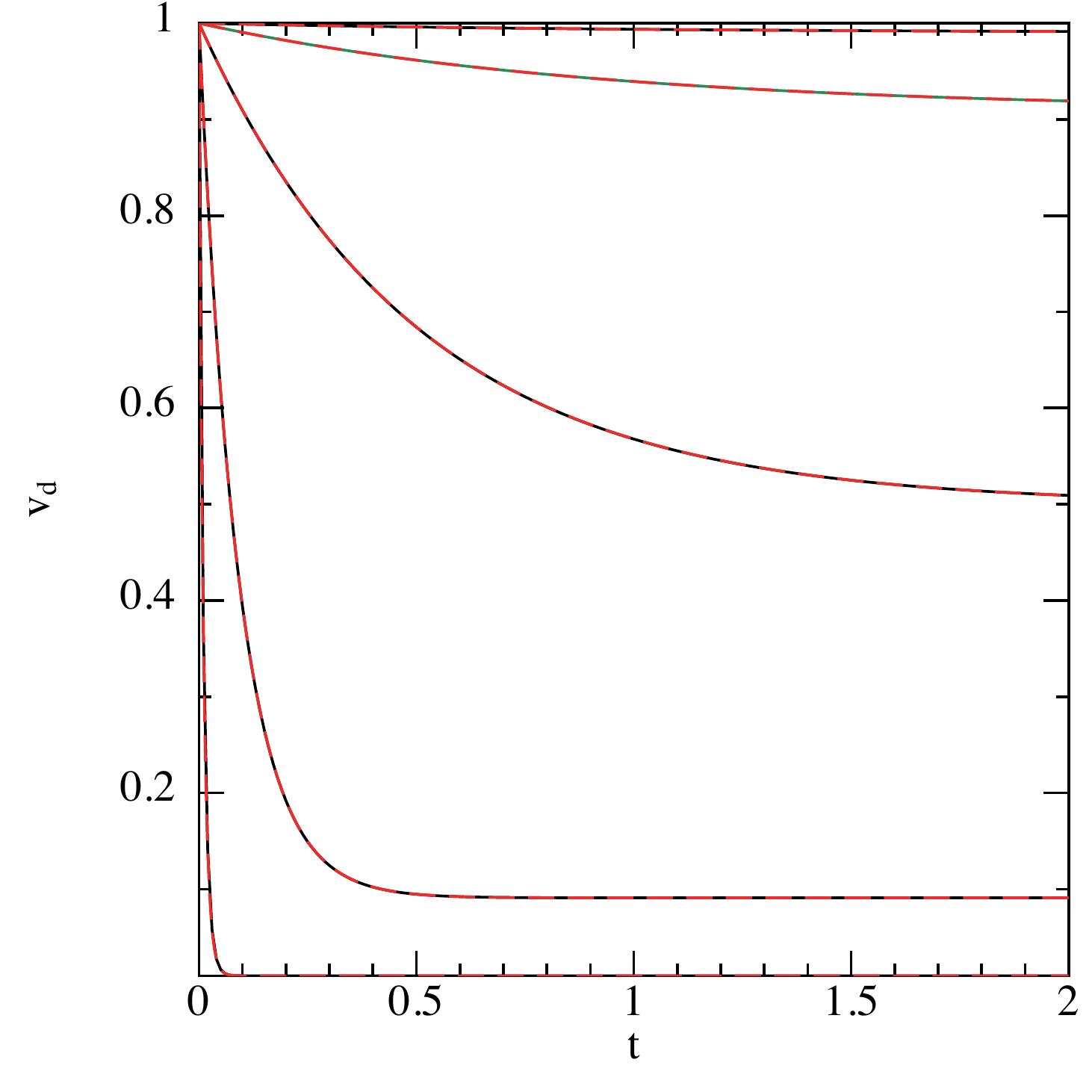}
   \caption{As in Fig.~\ref{fig:K} but varying the dust-to-gas ratio $\hrhod / \hrhod = 0.01, 0.1, 1, 10, 100$ (top-to-bottom, solid/black lines) with a fixed drag coefficient $K=1$. Exact solutions for each case are given by the long-dashed/red lines. Here again, the agreement is excellent (better than $1\%$).}
\label{fig:dtg}
\end{center}
\end{figure}

 Fig.~\ref{fig:K} shows the excellent agreement between the results obtained from the one-fluid algorithm (implicit integration) and the analytic solution of the \textsc{dustybox} problem ($< 1\%$ in the L1 norm) for a wide range of drag coefficients in the linear regime. In particular, this demonstrates the accuracy of the implicit algorithm developed in Sect.~\ref{sec:implicit}. For weak drag regimes ($\Delta t \ll \ts$), the simulation length is essentially the same as the one obtained using the explicit scheme, implying that the additional computational cost required to compute the implicit solution in the Predictor-Corrector scheme is negligible. However, for strong drag ($\ts\ll \Delta t$), the reduction in cpu time is of order $\Delta t / \ts$, improving the efficiency of the calculation significantly. Fig.~\ref{fig:dtg} shows that the one-fluid algorithm remains accurate for both small and large dust-to-gas ratios (or equivalently, dust fractions). Appendix~\ref{app:impquad} shows that similar results are also obtained for non-linear drag regimes.

\begin{figure*}
\begin{center}
   \includegraphics[width=\columnwidth]{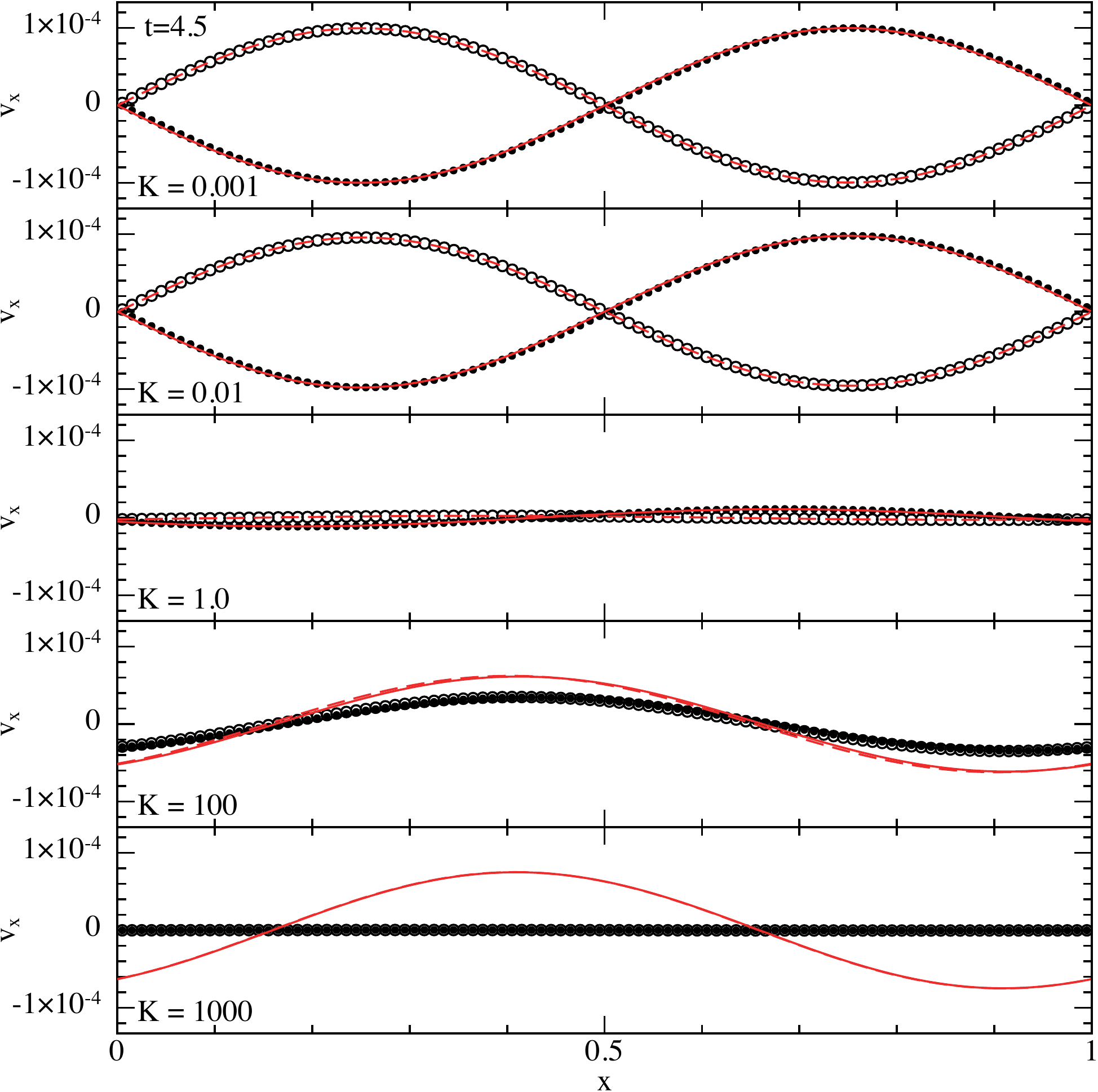}
   \hspace{0.5cm}
   \includegraphics[width=\columnwidth]{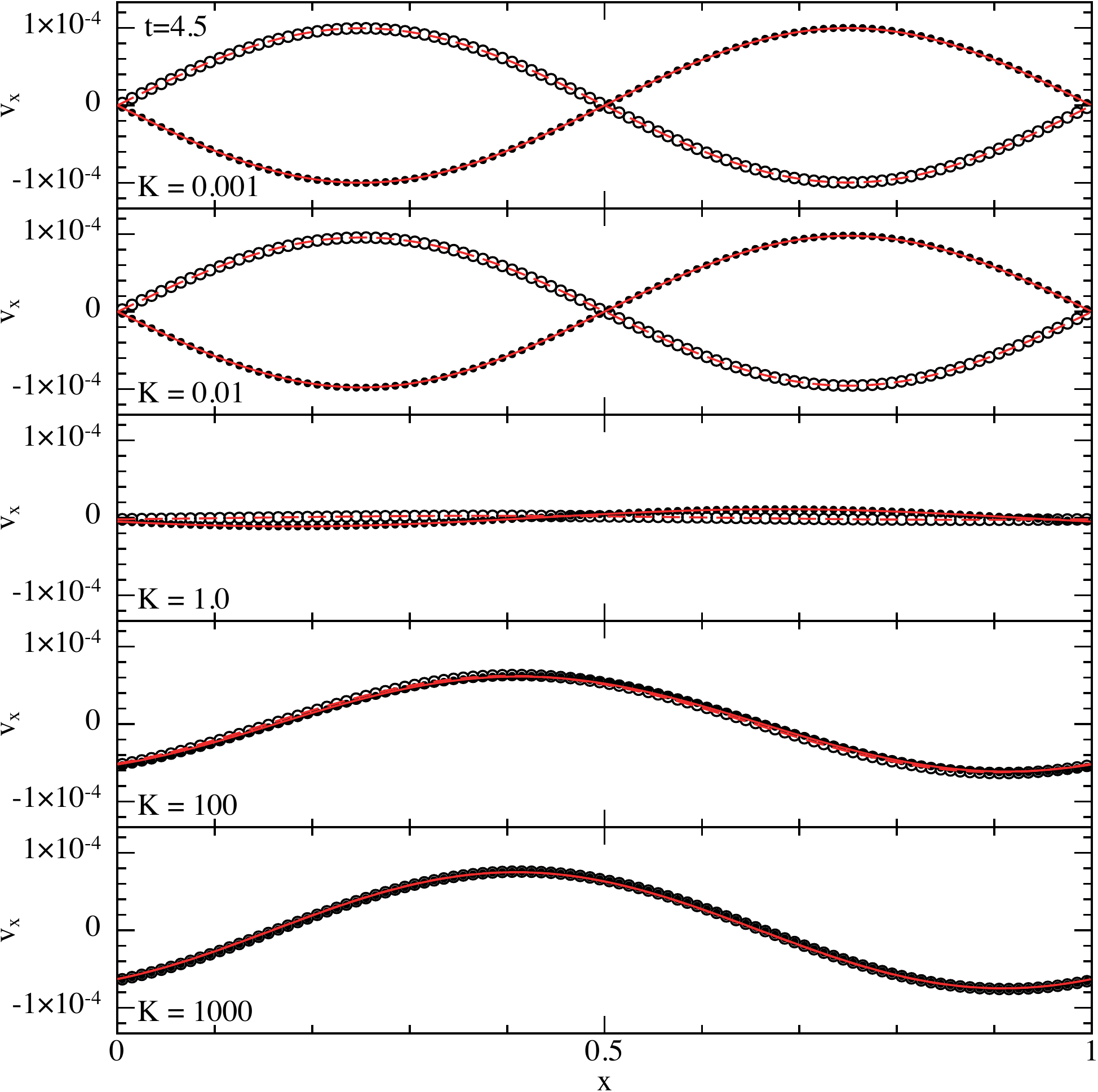} 
   \caption{The \textsc{dustywave} test, showing velocity of gas (filled circles) and dust (open circles) after 4.5 periods on the SPH particles with the two-fluid formulation (left) and the one-fluid formulation (right), compared to the analytic solution given by the solid and dashed red lines for the gas and dust, respectively. In the two fluid case the gas and dust are represented by $2 \times 100$ of particles, whereas in the one fluid formulation the two velocities are carried on the same set of 100 particles. The one fluid formulation solves the over-damping problem at low resolution present in the two fluid formulation at high drag (compare bottom two panels of each Figure). The slight phase error in the gas velocity in the two fluid formulation, caused by interpolation errors, is also not present in the one fluid case (compare top two panels of each Figure).}
   \label{fig:dustywave}
\end{center}
\end{figure*}

\subsection{\textsc{dustywave}}
\label{sec:dustywave}

The \textsc{dustywave} problem consists of linear sound waves propagating in a gas and dust mixture of uniform density and uniform dust fraction. The gas and dust phases interact via a linear drag term. The \textsc{dustywave} involves small perturbations, so it mainly tests the the treatment of forces specific to a single phase (here, the gas pressure) while neglecting the small contributions from the non-linear terms in the one-fluid formulation.

\subsubsection{Setup}

The equilibrium state is defined by the mixture at rest with $\deltav = 0$, implying that both the gas and the dust phases are at rest. We run 1D simulations with various combinations of the initial drag constant and dust fractions. SPH particles are distributed in the periodic domain $x \in [0,1]$ as described in \citet{LP12a}. No artificial viscosity is applied. We set the relative amplitude of the perturbation to $10^{-4}$ with respect to the quantities at equilibrium in order to remain in the linear acoustic regime for which the solution in \citet{LP11} is derived. We adopt an isothermal equation of state $P_{\rm g}= c_{\rm s}^{2} \rho_{\rm g}$ with $c_{\rm s} = 1$.

\citet{LP12a} showed that this problem proves difficult for two-fluid gas and dust codes when the drag coefficient is high. The main problem is that the small spatial dephasing between the gas and the dust generated by the gas pressure gradient must be resolved by the numerical algorithm. This was found to imply a drastic spatial resolution criterion $\Delta \lesssim c_{\rm s} t_{\rm s}$ (where $\Delta \equiv h$ in SPH) to avoid overdamping of the wave amplitude.

\subsubsection{Visualisation}
\label{sec:viz}
For the one-fluid algorithm, visualisation is less straightforward than the two-fluid version, since there is only one set of particles containing all of the information about both phases. The simplest way to visualise the mixture is to reconstruct two duplicate sets of particles, with the same positions and smoothing lengths, but with one set given the density, mass, velocity and internal energy of the gas:
\begin{align}
\rho_{\rm g, a} & = (1 - \epsd_{a}) \rho_{a}, \\
m_{\rm g, a} & = (1 - \epsd_{a}) m_{a}, \\
{\bf v}_{\rm g, a} & = {\bf v}_{a} - \epsd_{a} \deltav_{a}, \\
u_{a} & = u_{a},
\end{align}
and one set given the density, mass and velocity of the dust:
\begin{align}
\rho_{\rm d, a} & = \epsd_{a} \rho_{a}, \\
m_{\rm d, a} & = \epsd_{a} m_{a}, \\
{\bf v}_{\rm d, a} & = {\bf v}_{a} + (1 - \epsd_{a}) \deltav_{a}, \\
u_{a} & = 0.
\end{align}
This is the procedure we use in this paper, implemented in \textsc{splash} \citep{Price2007}, which enables a direct comparison to the two-fluid formulation.

\subsubsection{Results}
Fig.~\ref{fig:dustywave} compares the results obtained with the two-fluid algorithm and the one-fluid algorithm on the \textsc{dustywave} problem, in each case shown against the analytic solution derived in \citet{LP11} for both the gas and the dust phases. The drag coefficient is varied systematically from weak ($K = 0.001$) to strong ($K = 1000$) drag regimes. For the one-fluid simulations the number of SPH particles is fixed to $100$, with $2\times 100$ required for the two fluid calculations. The solutions of the \textsc{dustywave} problem can be seen to be well reproduced by the one-fluid algorithm. The accuracy is of order a few percent in the $L_{1}$ norm \textit{for every drag regime considered}, consistent with a second order integration scheme. Importantly, the direct comparison with the results obtained with the two-fluid algorithm (comparing left and right panels) shows that the spatial resolution criterion required for strong drag regimes is no longer needed. With $100$ SPH particles of each type per wavelength, the criterion $h < \cs \ts$ gives $K_{\rm l} \simeq 50$ as the maximum drag coefficient that can be simulated by the two fluid algorithm. The left panel of Fig.~\ref{fig:dustywave} shows that the wave amplitude is already incorrectly reproduced with the two-fluid method for $K = 100$ and that the wave in completely over-damped for $K = 1000$ (see Sect. 4.2 of \citet{LP12a} for a  quantitative discussion on the rate of energy over-dissipated in under-resolved simulations). To handle the case $K = 1000$, we would need to have used $2\times 2000  = 4000$ particles with the two-fluid algorithm. There is no resolution requirement in the one-fluid algorithm except the usual need to resolve a wavelength by $\sim 8$--10 particles, reducing the computational cost by a factor of $\sim 400$ in 1D. In 3D, the computational cost is reduced by a factor of $400^{3} = 64$ million: Accurate 3D simulations with the two-fluid algorithm at high drag would be inconceivable. An additional gain results from the fact that the implicit integration scheme for two-fluids converges slowly and is of limited utility when the drag is very strong \citep{LP12b}. An additional factor (of $\simeq 100$) is gained in the one-fluid algorithm from the ability to use an efficient implicit integration scheme, as described in Sect.~\ref{sec:implicit}. This implies a total improvement in speed of $6.4$ billion (this is not a misprint) in 3D.

\subsection{\textsc{dustyshock}}
\label{sec:dustyshock}

\begin{figure*}
\begin{center}
   \includegraphics[width=0.48\textwidth]{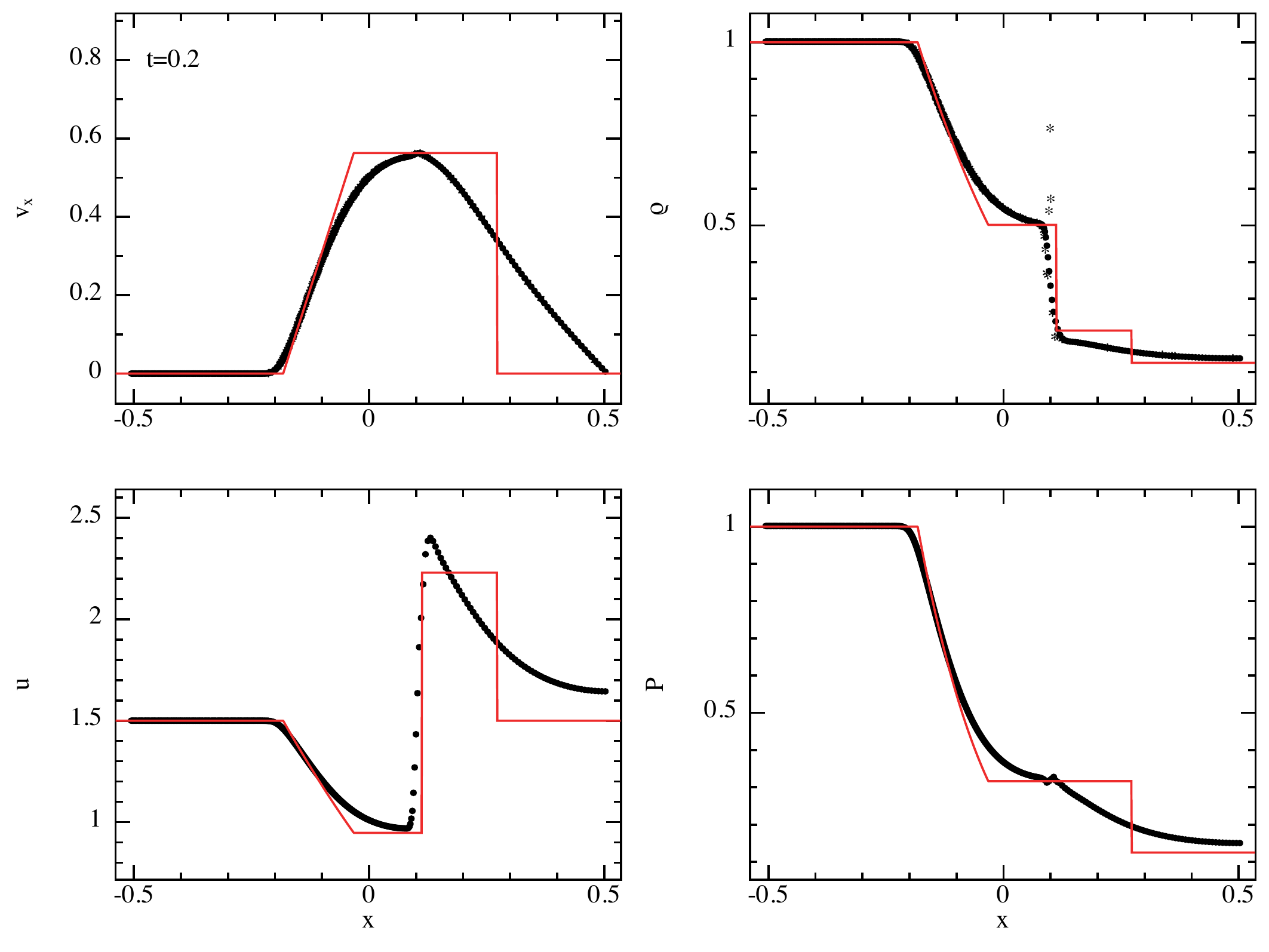}
   \hspace{0.02\textwidth} 
   \includegraphics[width=0.48\textwidth]{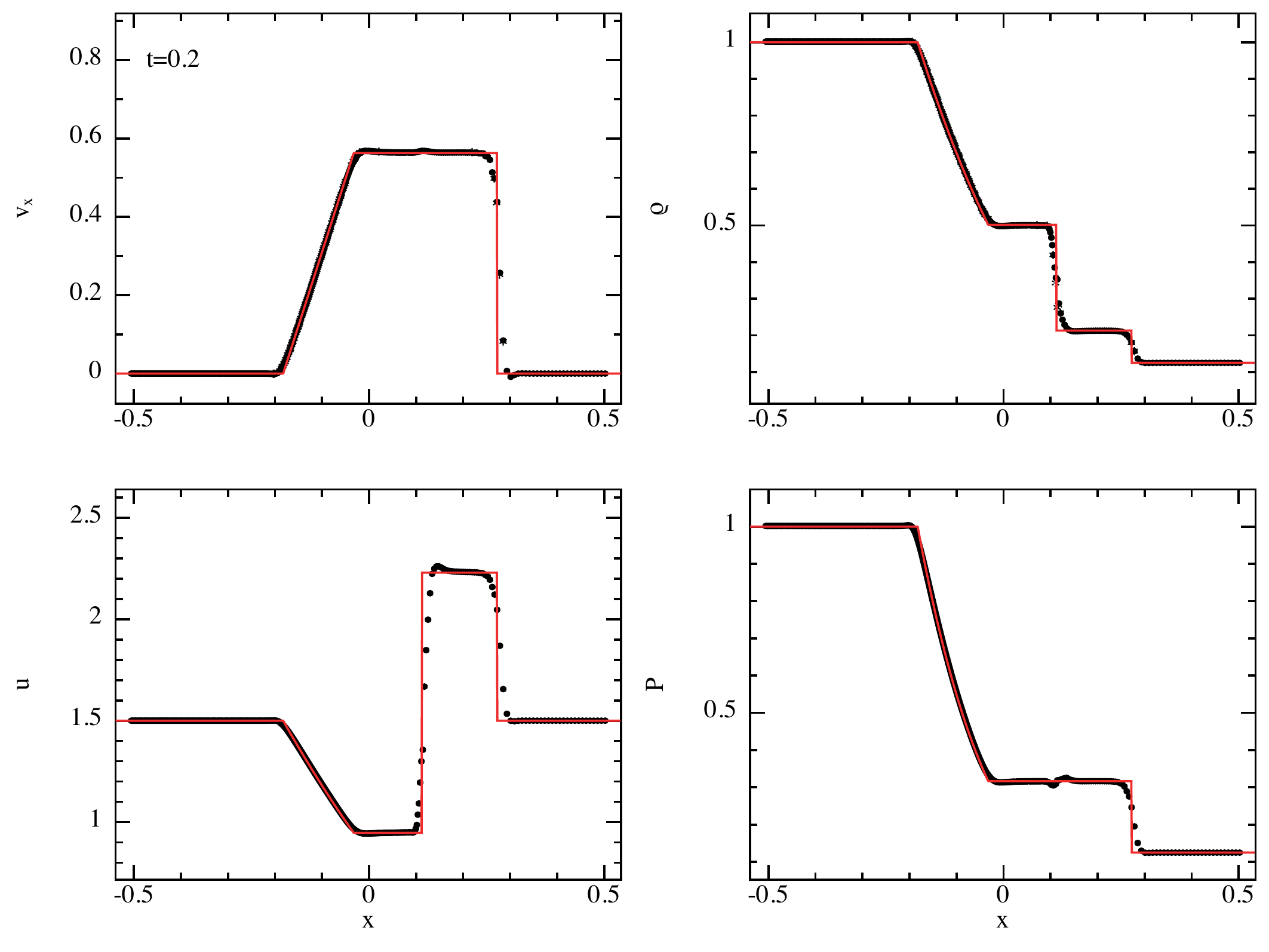} 
   \caption{Results of the \textsc{dustyshock} test with a high drag coefficient $K=1000$, comparing the two fluid formulation (left) and the one-fluid formulation (right) using 569 particles ($\times 2$ for the two fluid case) and the analytic solution for infinite drag given by the solid red line. Because the resolution criterion $h < c_{\rm s} t_{\rm s}$ is not satisfied, the two-fluid algorithm under-resolves the separation between the fluids and gives an incorrect solution (left panel). No such criterion is necessary with the one-fluid algorithm (right panel) and the correct solution (red line) is obtained.}
   \label{fig:dustyshock-strong}
\end{center}
\end{figure*}

The \textsc{dustyshock} problem involves the propagation of a shock in a dust and gas mixture. The problem is simplified by using a linear drag regime with constant drag term $K$, no heat transfer between the phases and no viscosity other than the standard shock-capturing terms used in SPH. After a transient phase, the shock is followed by a stationary phase that consists of the solution for a pure gas solution propagating at a modified $\gamma$ and sound speed, as described in \citet{LP12a}. In the \textsc{dustyshock} problem the advection of the mixture and the treatment of the discontinuity, which involves the SPH artificial viscosity terms, bring an additional complexity compared to the \textsc{dustybox} and the \textsc{dustywave} problems. As for the \textsc{dustywave} problem, the two-fluid dust and gas algorithm was found to poorly perform on the \textsc{dustyshock} problem for strong drag regimes since the spatial resolution criterion found by \citet{LP12a} has to be satisfied.

We have performed simulations of the \textsc{dustyshock} problem using $569$ particles, corresponding to a particle spacing of $\Delta x = 0.001$ for $x < 0$, varying the drag coefficient $K$ and the dust fraction.

\subsubsection{Strong coupling regime}

The right panel of Fig.~\ref{fig:dustyshock-strong} shows that the results obtained with the one-fluid algorithm for the \textsc{dustyshock} problem for strong drag regimes ($K = 1000$) are in excellent agreement with the analytic solution for perfect coupling in both the gas and the dust phases. The jump conditions for the velocity, the density, the internal energy and the gas pressure are reproduced with an accuracy of a few percent and the discontinuities are spread over 1--2 SPH smoothing lengths by the artificial viscosity and conductivity terms. In other words, the results are comparable to standard shock simulations with SPH \citep[e.g.][]{price12}. This confirms that the shock dissipation terms described in Sec.~\ref{sec:diss} are derived and implemented correctly.

 By contrast, using this number of particles ($\times 2$) with the two-fluid algorithm (left panel) means that the resolution criterion $h < \cs \ts$ is not satisfied, meaning that the small separation between the fluids is not resolved. This results in a numerical solution similar to what would be obtained with a physically smaller drag coefficient, still in the transient stage, which is incorrect. The comparison between the two methods (comparing left and right panels) shows that the large spatial resolution needed to avoid the over-dissipation of the kinetic energy in the shock is not necessary with the one-fluid algorithm (right panel), in agreement with our findings on the \textsc{dustywave} problem. This implies that a one-fluid algorithm is necessary to accurately simulate highly compressible systems involving strong drag.

\subsubsection{Weak coupling regime}

\begin{figure*}
\begin{center}
   \includegraphics[width=\columnwidth]{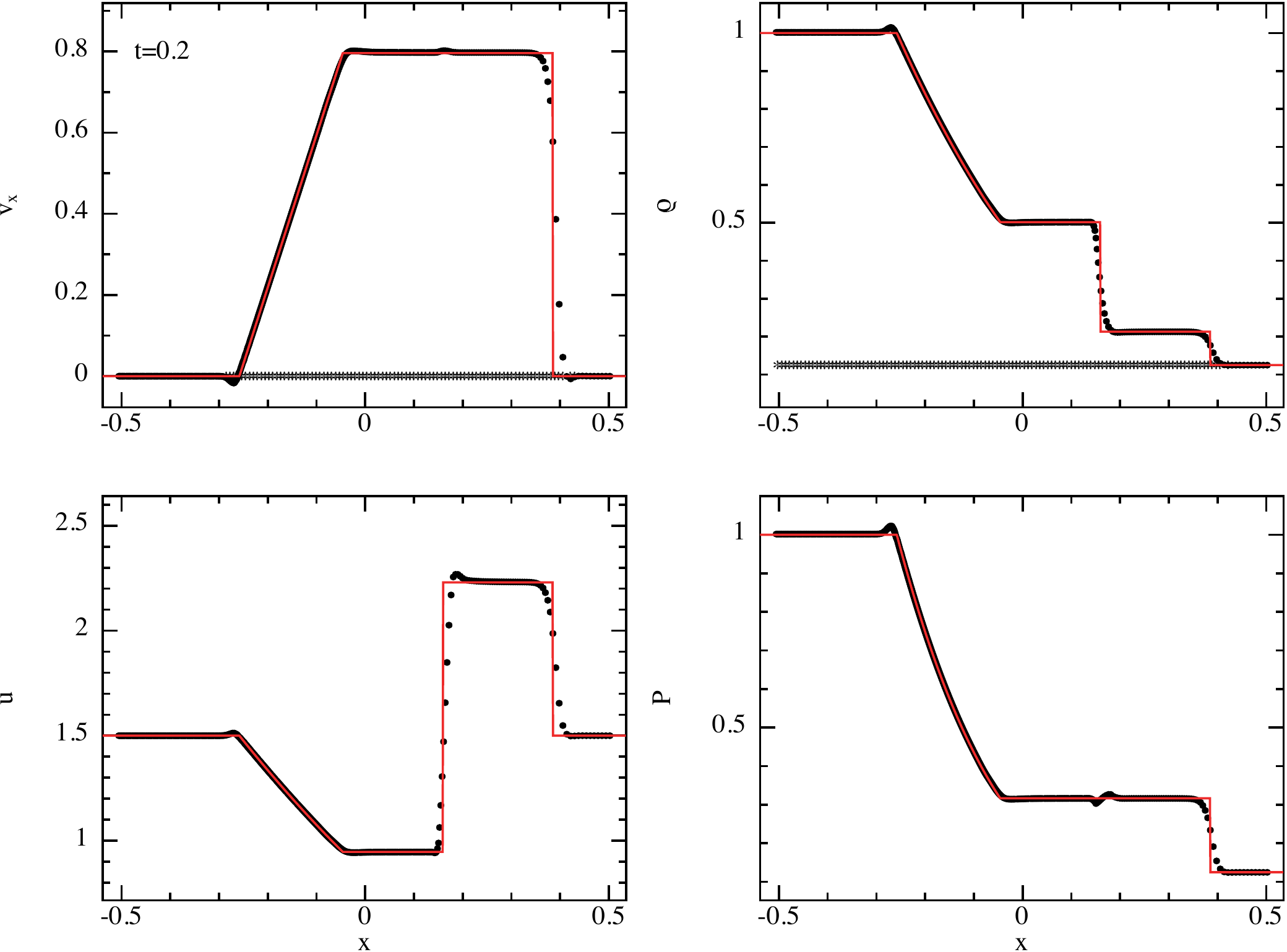}
   \hspace{0.5cm} 
   \includegraphics[width=\columnwidth]{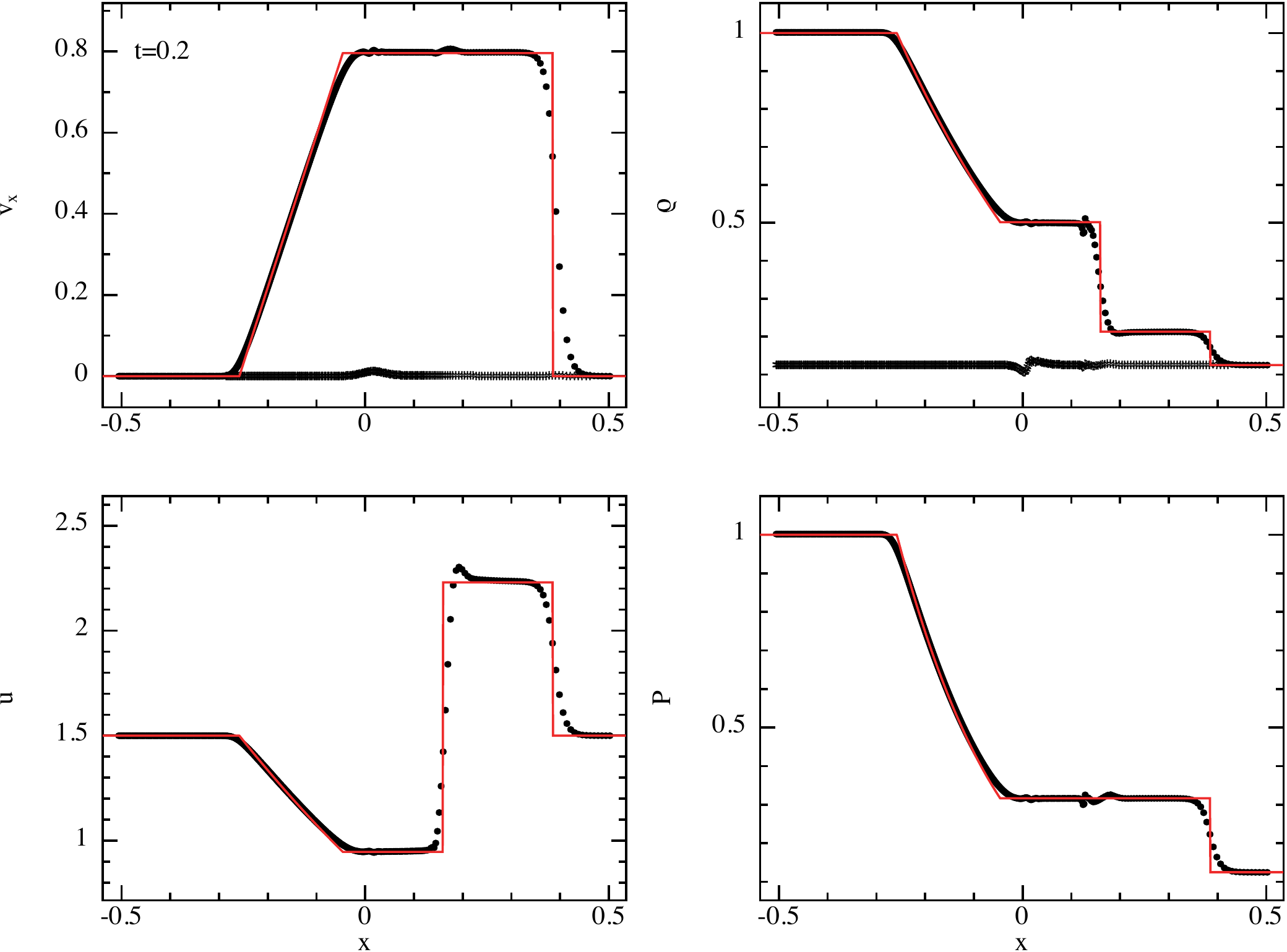} 
   \caption{Results of the \textsc{dustyshock} (Sod shock tube) test with zero drag with $\rhod = 0.1$ everywhere, comparing the solution with the two-fluid method (left) to the one-fluid method (right) and the analytic solution for the gas (red lines), using 569 particles (plus $126$ dust particles in the two-fluid case). The gas and dust velocities and densities are shown separately in the top row plots --- in the two-fluid case (left) these are the velocities and densities of the particles of each type, whereas in the one-fluid case (right) these are computed from the properties of the mixture. While the solutions are comparable, we have had to use a non-conservative formulation of the dissipation terms in the one-fluid case. Fig.~\ref{fig:dustyshock-alt} shows the solution when a conservative formulation of the dissipation terms is used.}
   \label{fig:dustyshock-weak}
\end{center}
\end{figure*}

 Fig.~\ref{fig:dustyshock-weak} shows the comparison between the two-fluid (left panel) and the one-fluid method (right panel) in the opposite extreme of no drag ($K=0$). This regime is trivial for the two-fluid algorithm: The gas follows the solution of the 1D Sod shock problem and the dust remains at rest.
The problem is much more difficult, even to conceptualise, with the one-fluid algorithm, requiring backwards advection of the dust fraction in order to keep the overall dust density unchanged as the shock propagates.

\begin{figure}
\begin{center}
   \includegraphics[width=\columnwidth]{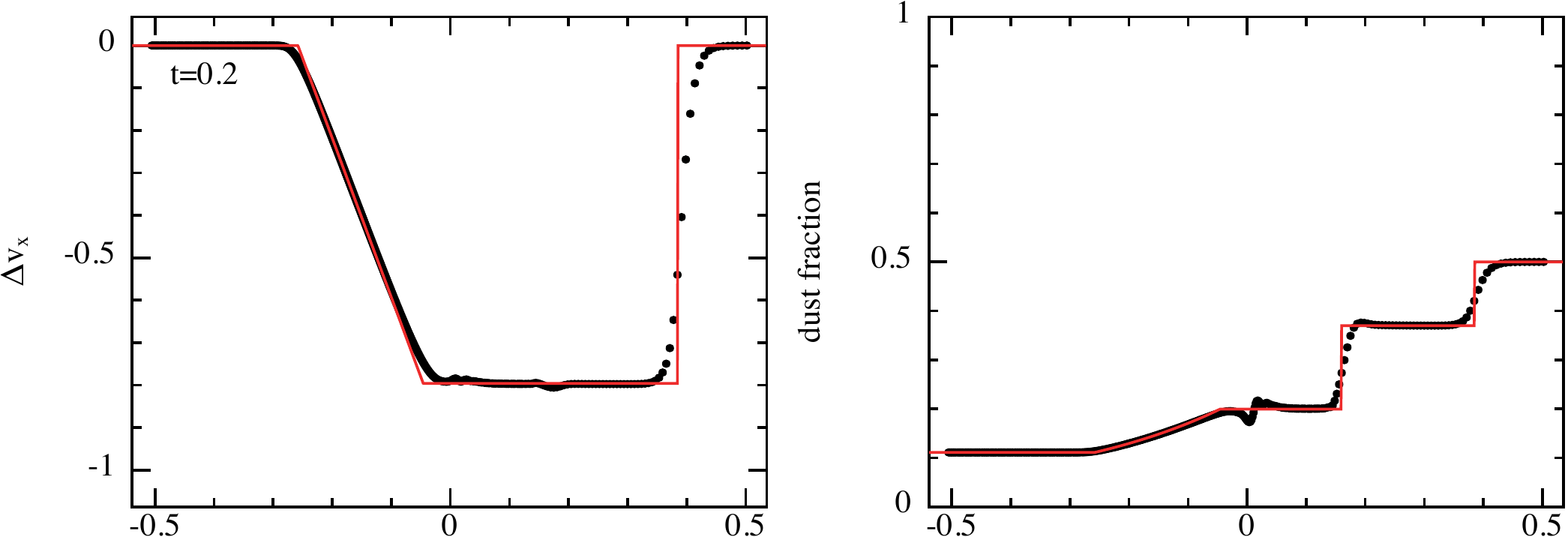}
   \caption{Evolution of $\deltav$ and $\epsilon$ for the \textsc{dustyshock} problem with zero drag, showing the backwards shock in the dust fraction and relative velocity necessary to keep the overall dust density constant. The analytic solution is given by the red line. }
\label{fig:deltav}
\end{center}
\end{figure}

\begin{figure}
\begin{center}
   \includegraphics[width=\columnwidth]{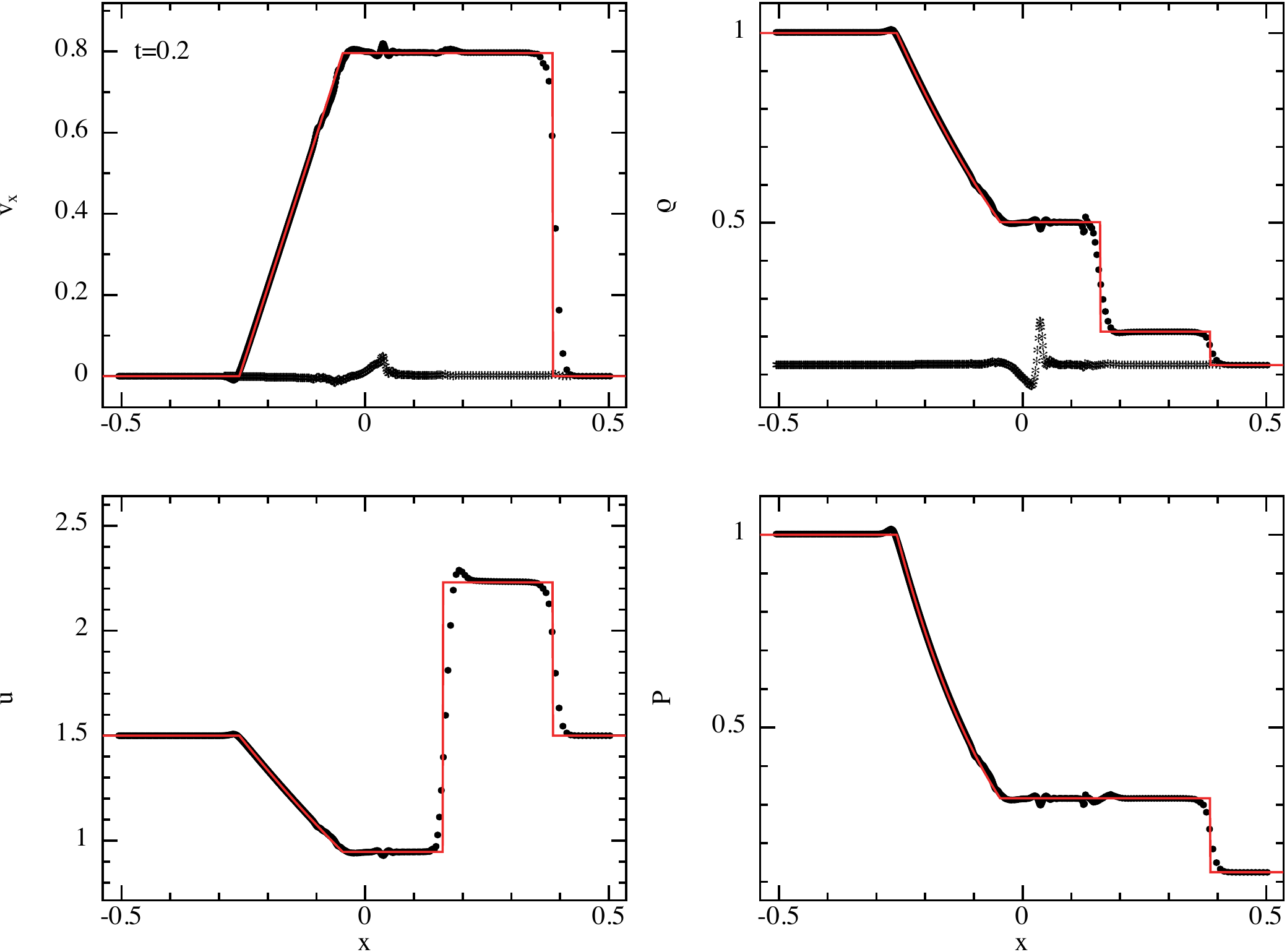} 
   \caption{As in Fig.~\ref{fig:dustyshock-weak} but using the conservative formulation of the dissipation terms in the one-fluid approach. In this case a post-shock blip in the dust density is found, but otherwise the solutions are similar.}
\label{fig:dustyshock-alt}
\end{center}
\end{figure}

 The right panel of Fig.~\ref{fig:dustyshock-weak} shows the solution obtained using the one-fluid method using the dissipation terms given in Sec.~\ref{sec:diss-nonconserv}. The jump conditions are well reproduced, but it was necessary to use the non-conservative formulation of the dissipation terms to eliminate post-shock oscillations in the backwards advection of the dust fraction (Fig.~\ref{fig:deltav} shows the evolution of $\deltav$ and $\epsilon$). Fig.~\ref{fig:dustyshock-alt} shows the best solution we could obtain using the conservative formulation of dissipation terms (as described in Sec.~\ref{sec:diss}; including the additional dissipation term in $\deltav$). In this case a post-shock `blip' occurs in the dust density, but otherwise the solution is similar. In the absence of the additional dissipation term (Eq.~\ref{eq:deltavdiss}) much stronger oscillations occur in both $\epsilon$ and $\deltav$, with a wavelength $\sim h$.
 
  In summary, while this problem is harder for the one-fluid algorithm compared to the trivial solution obtained with the two-fluid approach, it is possible to obtain a satisfactory solution for shocks in both the weak and strong coupling regimes.

\subsection{\textsc{dustyoscill}}
 Our final test, \textsc{dustyoscill}, is designed to probe the limitations of the one-fluid description of dust/gas mixtures in astrophysical problems. The problem is designed to mimic the settling of dust and gas to the midplane of an accretion disc. We consider a totally decoupled gas and dust mixture (i.e. no drag between the phases) evolving under the action of a linear force of the form $\mathbf{g} = - \Omega^2 \mathbf{z}$. The problem is one-dimensional along the vertical coordinate $z$, although we simulate it in two dimensions.  
 
\subsubsection{Problem description and analytic solution}
From a numerical point of view, the study is restricted to a box of size $2 \times \zmz$, centred on $z = 0$, where $\zmz$ is the initial height of the dust layer. Initially, both phases are at rest and the gas density is fixed to a constant $\rhog (z) = \rhogz$. Hydrostatic equilibrium in the gas is ensured by setting the gas sound speed, $\cs$, according to
\begin{equation}
\cs^{2} \left( z \right) = \cs^{2}\left( \zmz \right) -\frac{\Omega^{2}}{\rhogz} \left(\zmz ^{2}  - z^{2} \right) .
\label{eq:verteq}
\end{equation}
This differs from the physical case of the midplane in a protoplanetary disc, since it avoids unnecessary complications due to gas stratification. The dust phase consists of a layer of homogeneous density $\rhodz$ distributed over $\left( \left| z \right| \le  \zmz  \right)$ and initially at rest. 

The two phases quickly separate as the system evolves --- in the absence of coupling with the dust, the gas remains remains at hydrostatic equilibrium, whereas the dust particles should oscillate with frequency $\Omega$ around $z = 0$. More precisely, the position $\zdl$ and the Lagrangian velocity $\vdl$ of a dust particle initially located at $z_{\rm 0}$ are given by:
\begin{eqnarray}
\zdl  \left( t \right) & = & z_{\mathrm{0}} \cos \left(\Omega t \right),\\
\vdl \left( t \right) & = & - \Omega z_{\mathrm{0}} \sin \left(\Omega t \right).
\label{eq:lagrang}
\end{eqnarray}
This problem exhibits an interesting mathematical singularity since all the particles cross $z = 0$ at the same time, meaning that the dust density becomes infinite at this moment. This property is specific to the action of a linear force, well-known for generating isoschronic oscillations. In general, the oscillation period would have depended on the initial position of the particle. The main point is that, in this general case, once the first particles have crossed $z = 0$, the velocity field becomes locally multi-valued and the dust phase can no longer be described as a fluid. Hence, this can be used to demonstrate an intrinsic limitation of the one-fluid approach.

The half-thickness of the dust layer $\zm$ is given by
\begin{equation}
\zm \left( t \right) =  \zmz \left|  \cos \left(\Omega t \right) \right| .
\label{eq:half_bla}
\end{equation}
Conservation of the total dust mass
\begin{equation}
\zm \left( t \right) \rhod \left( t \right) = \zmz \rhodz,
\label{eq:total_mass}
\end{equation}
implies that the evolution of the dust density is given by
\begin{equation}
\rhod \left( t \right) = \frac{\rhod \left( t = 0 \right)}{ \displaystyle \left| \left( \frac{\partial \zdl}{\partial z_{\mathrm{0}}} \right)_{t} \right| } = \frac{\rhodz}{\left| \cos \left(\Omega t \right) \right|} ,
\label{eq:rhodt}
\end{equation}
where the absolute value on the denominator is required since the dust particles cross the $z=0$ plane every half-period.%

The equations of mass and momentum conservation for the dust phase are:
\begin{eqnarray}
\frac{\partial \rhod}{\partial t} + \frac{\partial \rhod \vdz}{\partial z}   & = & 0, \label{eq:dust_mass}\\
\frac{\partial \vdz }{\partial t} + \vdz \frac{\partial \vdz }{\partial z}  & = & - \Omega^2 z, \label{eq:dust_momentum}
\end{eqnarray}
where $\vdz$ is the Eulerian velocity of the dust phase. With the dust density given by Eq.~\ref{eq:rhodt}, the analytic solution of Eqs.~\ref{eq:dust_mass} -- \ref{eq:dust_momentum}  is
\begin{equation}
\vdz =
\begin{cases}
 - \Omega z \tan \left( \Omega t \right) , & \left| z \right| \le \zm \left( t \right) ; \\
 0,& \left| z \right| > \zm \left( t \right) .
\end{cases}
\label{eq:vd_dust}
\end{equation}

\subsubsection{Challenges}
The \textsc{dustyoscill} problem is challenging for our one-fluid dust/gas description, for three reasons:
\begin{enumerate}
\item The two phases evolve independently, which is not well described as a mixture. The limit is one of very weak coupling that may be encountered in astrophysical systems. 

\item The moving boundary $z = \pm \zm \left( t \right)$ represents a discontinuity in the dust density and hence the dust fraction of the mixture. For $\left| z \right| \le \zm \left( t \right)$, the dust density is non zero as indicated by Eq.~\ref{eq:rhodt}. However, for $\left| z \right| > \zm \left( t \right)$, $\rhod = 0$ exactly. The numerical method must model this discontinuity correctly in order to avoid negative values of $\rhod$. 

\item A singularity occurs, and the dust velocity field becomes multi-valued, when the particles cross $z= 0$, i.e. when $\Omega t = n \pi / 2$, for integer $n$. This is not a problem when the mixture particles represent dust alone --- our method in that case reverts to the usual particle description of dust --- but with some gas in the mixture, we expect a multi-valued velocity field to be problematic.
\end{enumerate}

\subsubsection{Setup}

We set up the \textsc{dustyoscill} problem using $80 \times 20$ particles in 2D, distributed on a square lattice in the domain $x \in [-2,2]$ and $z \in [-0.5,0.5]$ in code units. The boundary conditions are periodic along the $x$ axis and free along the $z$ axis. No artificial viscosity is applied. Initially the total density is uniform, $\rho = 1$. The dust-to-gas ratio is set to $\rhod / \rhog = 0.01$, implying a dust fraction $\epsd = 1 / 101$. The gas sound speed at $z = 0$ is set to $c_{\rm s, max} = 5$, such that the travel time of a sound wave is smaller than the oscillation period. We use $\zmz = 1$ and $\Omega = 1$ in code units. We also performed simulations with the two-fluid method described in \citet{LP12a} for comparison. In this case, twice as many particles were used, with the dust particles initially superimposed on the gas particles.

\subsubsection{Results}

%\begin{figure}
%   \includegraphics[width=\columnwidth]{Figures/dustyoscill.pdf}
%   \caption{TVelocities (top panel) and densities (bottom panel) are given as a function of the coordinate $z$. The agreement between the two simulations is quite good. The two-fluid algorithm picks better the discontinuity in the dust density whereas it is spread over a few smoothing lengths in the one-fluid case. However, ten times less particles are required to handle the dust concentration in the one-fluid case. Non zero dust velocities obtained at locations where $\rhod = 0$ are just an artefact of the visualisation post-treatment. \todo{split this Figure into left and right panels showing two fluid and one fluid}}
%   \label{fig:dustyoscill}
%\end{figure}

\begin{figure}
\begin{center}
   \includegraphics[width=0.49\columnwidth]{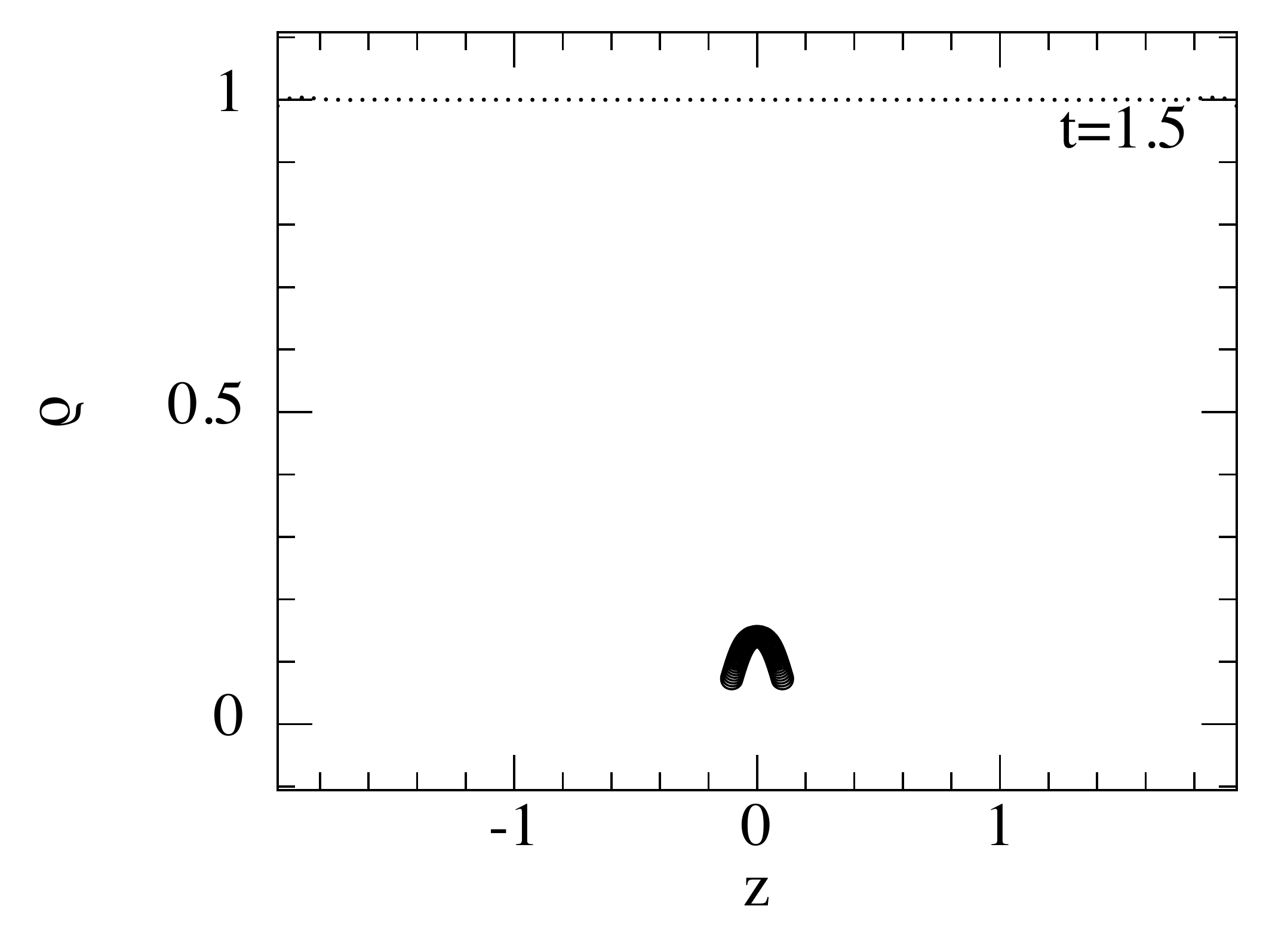}
   \includegraphics[width=0.49\columnwidth]{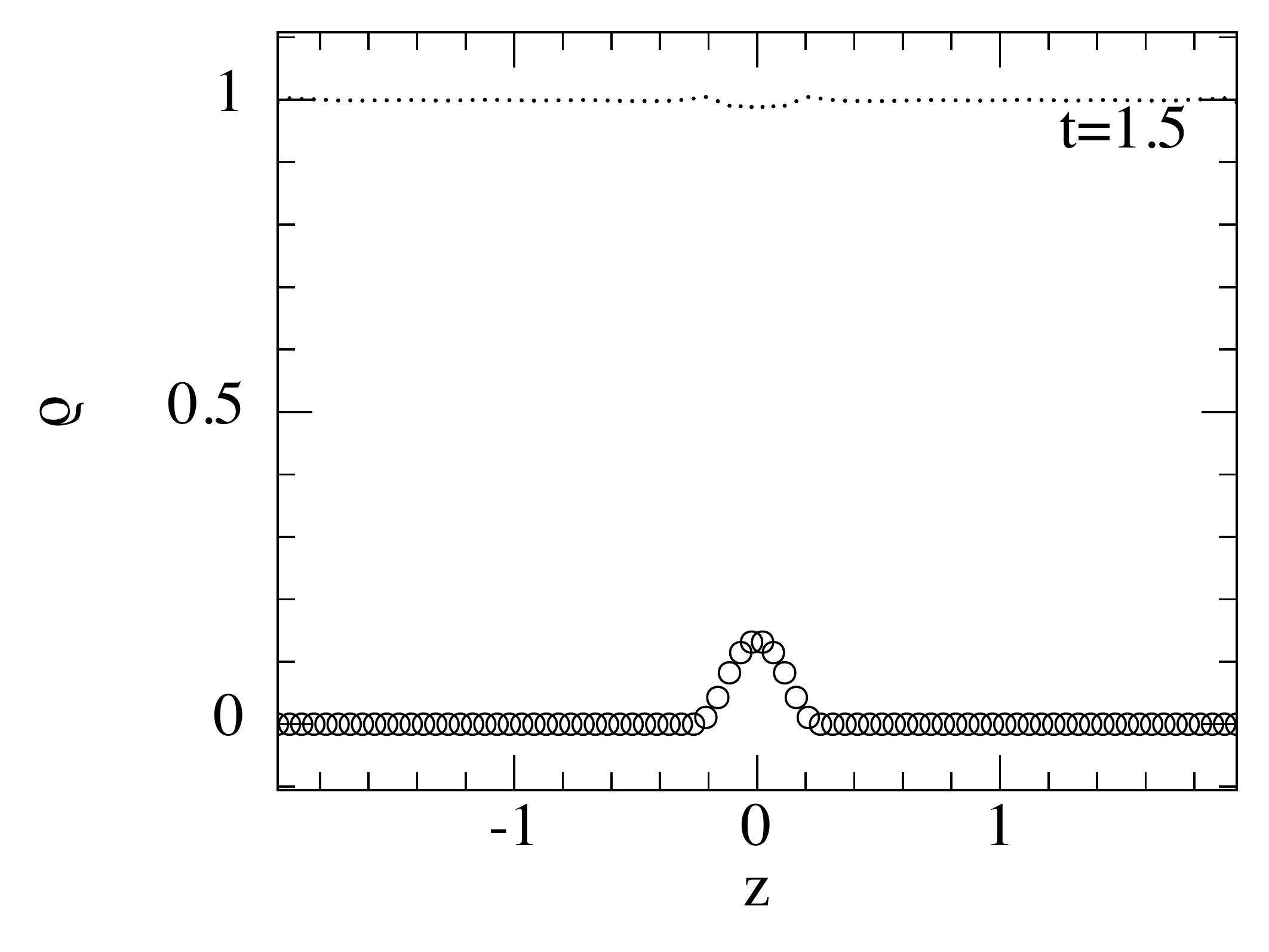} 
   \caption{Density as a function of $z$ in the \textsc{dustyoscill} problem at $t = 1.5$ (in code units) as treated by the two-fluid algorithm (left) and the one-fluid algorithm (right). Black dots and circles represent gas and dust particles (see Sec.~\ref{sec:viz} for the visualisation procedure in the one-fluid case). The two methods were found to produce similar results at this stage. The discontinuity in the dust density its slightly sharper with the two-fluid algorithm. On the other hand, 10$\times$ fewer particles are required to capture the dust concentration with the one-fluid technique.}
   \label{fig:dustyoscill}
\end{center}
\end{figure}

\begin{figure}
\begin{center}
   \includegraphics[width=0.49\columnwidth]{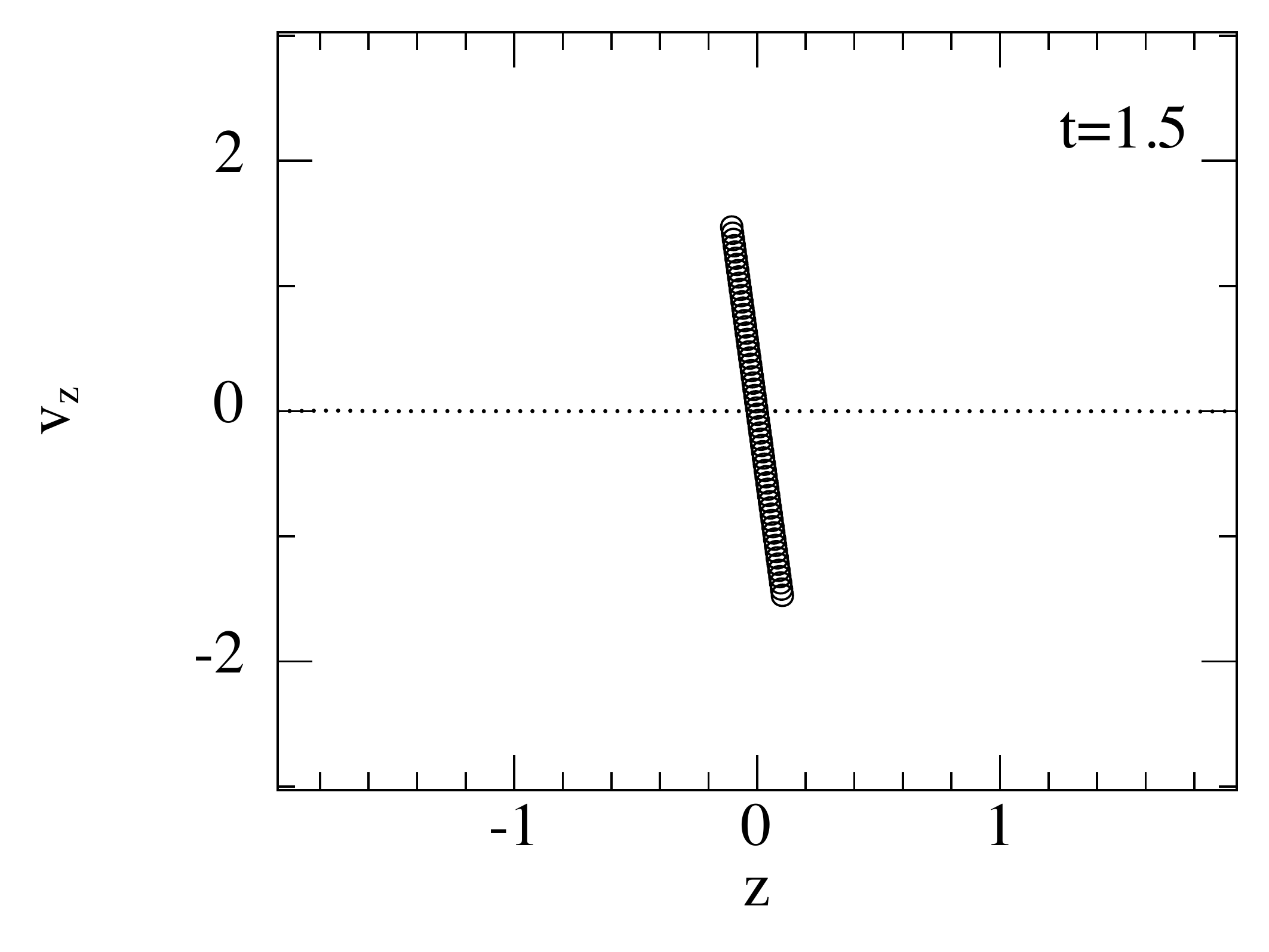}
   \includegraphics[width=0.49\columnwidth]{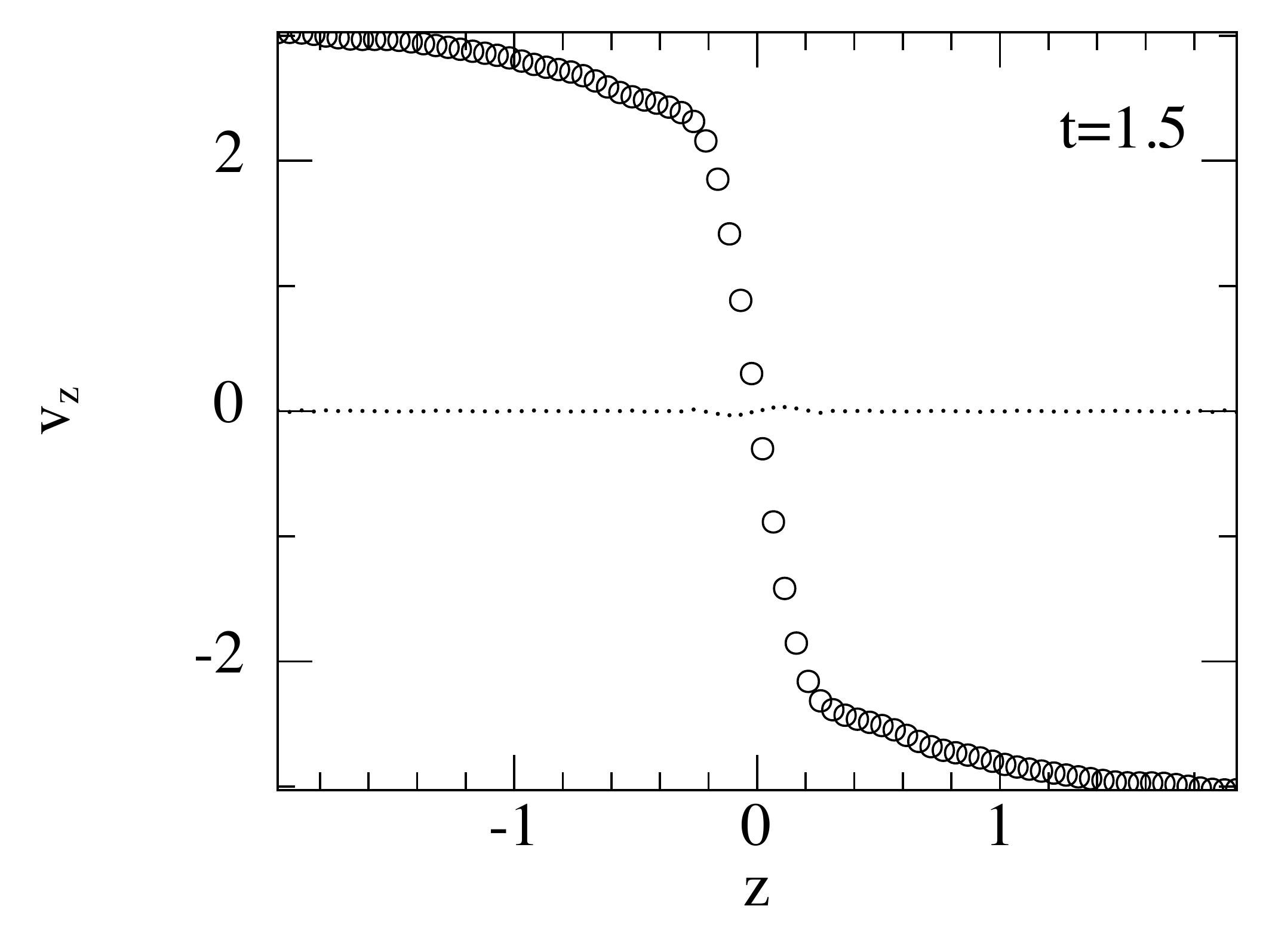} 
   \caption{Phase space diagram showing velocity as a function of $z$ for the problem shown in Fig.~\ref{fig:dustyoscill}, with the two-fluid (left) and one-fluid (right) methods. This shows the state of the simulations just before the dust layer crosses the origin and the dust velocity field becomes multi-valued (Fig.~\ref{fig:cross}).}
   \label{fig:phase}
\end{center}
\end{figure}

 This problem, by construction, is trivial with the two-fluid formulation, since the two phases of the mixture are decoupled and independent. As such, SPH simulations using the two fluid formulation show perfect agreement with the analytic solution of the problem (left panel of Fig.~\ref{fig:dustyoscill}): the gas remains at hydrostatic equilibrium during the simulation (with the exception of a small jitter observed close to the boundaries) and the dust particles experience harmonic motion with a period $\pi$ around $z = 0$. In the phase space $(z(t),v_{z}(t))$ (Fig.~\ref{fig:phase}), the gas particles lie on an horizontal static line, whereas the dust particles form a line which rotates clockwise at a uniform angular speed $\Omega$. 
 
\begin{figure}
\begin{center}
   \includegraphics[width=0.49\columnwidth]{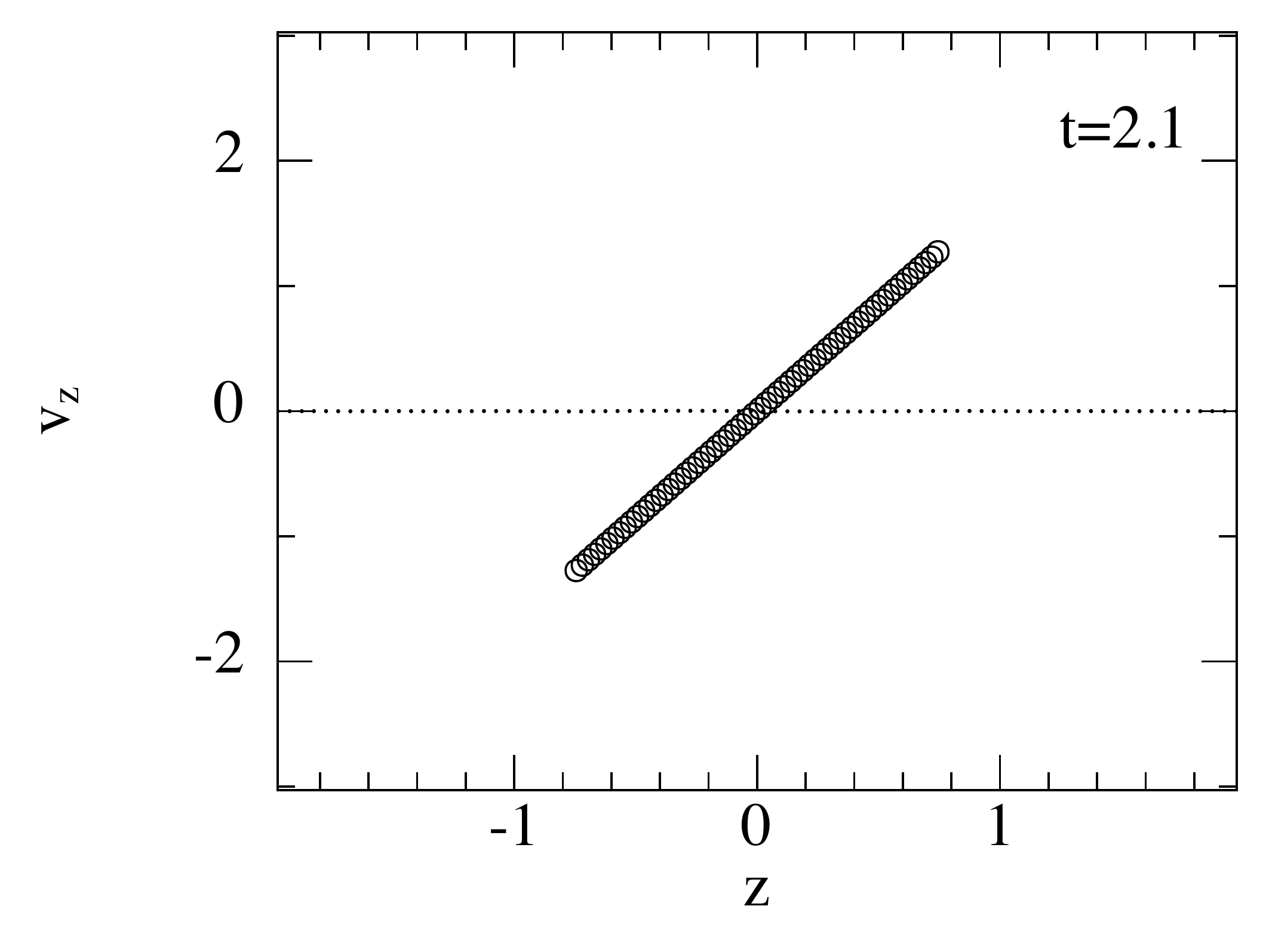}
   \includegraphics[width=0.49\columnwidth]{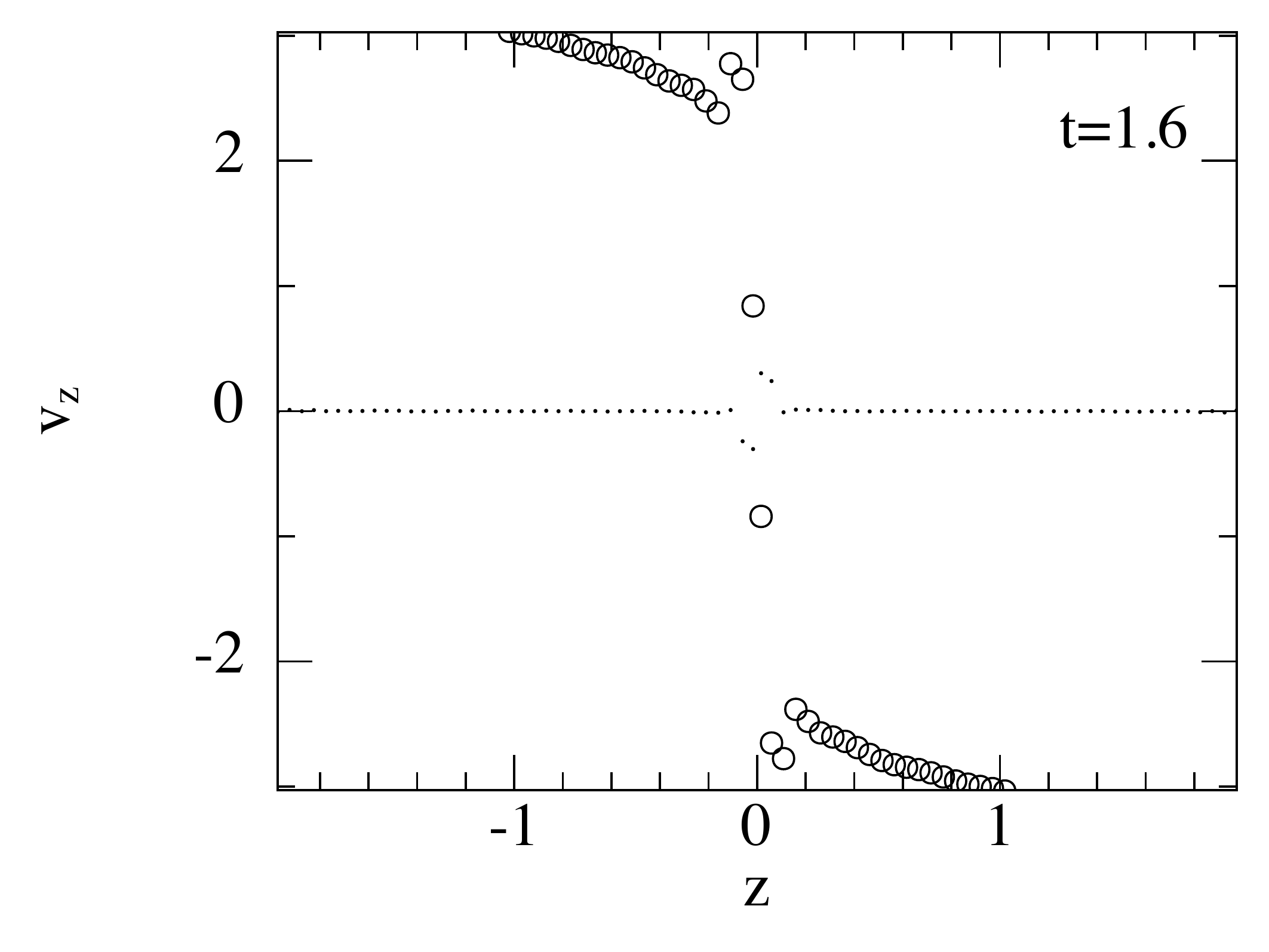} 
   \caption{As in Fig.~\ref{fig:phase}, but at a slightly later time, after the dust particles cross $z = 0$ for the first time (i.e. $t = \pi / 2$). The two-fluid algorithm captures the singular behaviour without difficulty (the left panel shows the state of the system at $t = 2.1$). The one-fluid algorithm, however, crashes at the singularity, suggesting that the one fluid description breaks down once the velocity field becomes multi-valued.}
   \label{fig:cross}
\end{center}
\end{figure}

 The first mathematical singularity occurs for $t = \pi/2$ (code units), when particles cross $z = 0$ for the first time. Again, the two fluid formulation picks up this singularity accurately (left panel of Fig.~\ref{fig:cross}): mass conservation is preserved regardless of the method used to compute SPH densities --- whether by direct summation over the masses of the neighbours (Eq.~\ref{eq:rhosum}) or via the SPH-discretised continuity equation. This shows that the method of evaluating the density, common to both the two-fluid and one-fluid formulations, is able to handle the discontinuity in the rate of change of density $\mathrm{d}\rho/\mathrm{d}t$ that is involved in the \textsc{dustyoscill} problem.

 The right panel of Fig.~\ref{fig:dustyoscill} shows the results using the one-fluid algorithm (as previously, the gas and the dust velocities in the plots are reconstructed from the properties of the mixture particles, as described in Sec.~\ref{sec:viz}). In the one-fluid simulation, the mixture particles are almost at rest since the gas --- which represent most of the mixture's mass --- remains in hydrostatic equilibrium. The simulation can be divided in two stages: the first one where the dust reaches $z = 0$ (typically $t \lesssim 1.5$) and a second stage where the dust effectively crosses $z = 0$ ($1.5 \lesssim t < \pi/2$). 

Fig.~\ref{fig:dustyoscill} shows that both the phase space profile and the densities are reproduced accurately by the one-fluid algorithm. The discontinuity in the dust density at $\zm$ is better picked-up by the two-fluid algorithm. Indeed, this discontinuity is spread over a few smoothing lengths $h$ by the one-fluid algorithm. However, fewer particles are required to handle the dust over-concentration in the one-fluid algorithm ($8$ instead of $80$ at $t = 1.5$). It should be noted that close to the discontinuity $z \simeq \zm(t)$, the dust fraction may become slightly negative (of order $10^{-5}$) because of SPH interpolation errors (this effect is reduced by using a better kernel, e.g. $M_{6}$ instead of $M_{4}$). We did not find any spurious effect due to this small negative value. Using a limiter (e.g. fixing $\epsilon = 0$ when $\epsilon < 10^{-6}$) was not found to change the results. However, this spurious effect should be kept in mind. Since $\epsilon$ is not exactly zero, our post-treatment will compute a value for the dust velocity which is also not zero, as seen in Fig.~\ref{fig:phase}.

However, the algorithm is not able to capture the singularity that arises when particles cross $z = 0$: the simulation breaks at $t = \pi/2$. When reaching this discontinuity, the slope of straight line $\left(z,v_{z} \right)$ in the phase-space diagram steepens as expected. However, this steepening saturates, providing an inaccurate simulation of the mixture at the discontinuity.

 To understand the origin of this limitation, we have analysed the role played by terms specific to the one-fluid formalism in Appendix~\ref{app:dustyoscill_theory}. The main cause appears to be that the last two terms of the right-hand side of Eq.~\ref{eq:genmomentum_deltav} do not counterbalance each other as expected, giving rise to an infinite rate of change for the differential velocity, which breaks the simulation. While spectacular, it should in kept in mind that this is the only problem for which we managed to break the one-fluid algorithm, which corresponds to the physical situation of a multi-valued velocity field in the dust, where the fluid description breaks down. In other problems (\textsc{dustywave} in particular), the one-fluid algorithm gives accurate results even when the phases are totally independent, despite the fact it was designed mainly to solve the fundamental limitations of the two-fluid algorithm in dealing with well-coupled mixtures.

%=======================================================================================================
\section{Conclusion}
\label{sec:conclu}

We have developed an SPH one-fluid gas and dust algorithm that is fully conservative. The gas and dust are treated as the two phases of the same fluid, the dust-and-gas mixture. This fluid has the total mass of its two phases and is advected at the barycentric velocity. The dust fraction and the differential velocities between the two phases are internal properties of the mixture. The resulting algorithm has been benchmarked against a suit of problems including \textsc{dustybox}, \textsc{dustywave}, \textsc{dustyshock} and \textsc{dustyoscill}.

The one-fluid algorithm was found to perform particularly well in the situation for which it was designed --- strong drag --- i.e. when dealing with the dynamics of small dust grains. In this situation it was found to be both more efficient (by a factor of $\sim 10^{10}$!) and more accurate than the standard approach of two fluids coupled by a drag term. Specific advantages compared to the two-fluid approach were found to be as follows:
\begin{enumerate}
\item The spatial resolution criterion $h < \cs\ts$ found by \citet{LP12a} for two-fluid simulations is no longer required;
\item Implicit time integration can be easily implemented and the resulting scheme is very efficient;
\item It avoids resolution issues when dust over-concentrates with respect to the gas, since there is only one resolution length in the mixture;
\item Interpolation errors caused by interpolation between two separate sets of particles are avoided;
\item A factor of two fewer particles are needed;
\item Only one smoothing length is required;
\item The standard bell-shaped SPH kernel is used throughout;
\item The formalism can be implemented in a hydro code without requiring special architecture for multiple fluid species.
\end{enumerate}

The main limitation to the one-fluid method was that the method breaks when the fluid approximation breaks down, i.e. when the velocity field becomes multi-valued.
For the most general case where simulations involve mixtures of both large grains and small grains, this suggests that a hybrid scheme might be the best approach. This would be straightforward to implement, consisting of a one-fluid mixture for the small grains, coupled to a separate large grain population treated via the usual two-fluid approach.

The main conclusion of this paper is that the one-fluid algorithm makes possible accurate numerical treatment of strong drag/small grains by removing the fundamental limitations of the two-fluid approach in this regime. Extensions of this study will involve the treatment of a distribution of grain sizes (at the moment, only a single grain size is considered), as well as applications to astrophysical systems where the role of small grains are dominant.

\section*{Acknowledgments}
We thank Sarah Maddison, Ben Ayliffe and Tom Hendricks for useful discussions. We also thank the anonymous referee for comments which have improved this paper substantially. This work was funded via Australian Research Council (ARC) Discovery Project grant DP1094585. G. Laibe acknowledges funding from the European Research Council for the FP7 ERC advanced grant project ECOGAL. DJP is very grateful for funding via an ARC Future Fellowship (FT130100034).

\bibliography{dustSPH}

\begin{appendix}

\begin{onecolumn}
\section{Limiting behaviour of the one-fluid equations for the \textsc{dustyoscill} problem}
\label{app:dustyoscill_theory}
 In this appendix, we analyse the limiting behaviour of the one-fluid equations for the \textsc{dustyoscill} problem, as the dust layer reaches $z=0$ for the first time, in order to understand the breakdown of the single fluid description for this problem. The analytic expressions of the comoving derivatives of $\rho$, $\rhod / \rho$, $\mathbf{v}$ and $\Delta \mathbf{v}$ are given by Eqs.~\ref{eq:genmass_rho} -- \ref{eq:genmomentum_deltav}. They highlight the evolution of the physical quantities when the particles reach $z = 0$ for the first time:

%\begin{multicols}
\begin{align}
 \lim\limits_{\substack{\Omega t \to \pi / 2^{-}}}  \left\lbrace \frac{\mathrm{d} \rho}{\mathrm{d} t}  = \rhodz \Omega \frac{\displaystyle \tan \left( \Omega t \right)}{\displaystyle  \cos \left( \Omega t \right)} \right\rbrace  & = +\infty, \label{eq:sys1}\\
 \lim\limits_{\substack{\Omega t \to \pi / 2^{-}}}  \left\lbrace  \frac{\mathrm{d} }{\mathrm{d} t} \left( \frac{\rhod}{\rho} \right) =  \frac{\rhodz \rhogz \Omega \sin \left(\Omega t \right)}{\left( \rhogz \cos \left( \Omega t \right) + \rhodz \right)^{2}} \right\rbrace & = \frac{\rhogz}{\rhodz} \Omega, \label{eq:sys2}\\
 \lim\limits_{\substack{\Omega t \to \pi / 2^{-}}}  \left\lbrace \frac{\mathrm{d} \mathbf{v} }{\mathrm{d} t} = \frac{\rhodz \left(\rhogz \cos \left(\Omega t \right)^{2} - \rhodz \cos \left(\Omega t \right) - 2 \rhogz  \right)\Omega^{2} z}{\cos \left(\Omega t \right)\left( \rhogz \cos \left(\Omega t \right) + \rhodz  \right)^{2}} \right\rbrace &  = \displaystyle \frac{-2 \rhogz \Omega^{2} \zm}{\rhodz} , \label{eq:sys3}\\
 \lim\limits_{\substack{\Omega t \to \pi / 2^{-}}}  \left\lbrace   - \frac{1}{\rho}\nabla\cdot \left(\frac{\rhog \rhod}{\rho} \deltav \deltav \right) = -2 \zm \Omega^{2} \frac{\sin \left(\Omega t \right)^{2}}{\rhodz} \right\rbrace &  = \displaystyle \frac{-2 \rhogz \Omega^{2} \zm}{\rhodz} , \label{eq:sys4}\\ 
 \lim\limits_{\substack{\Omega t \to \pi / 2^{-}}}  \left\lbrace  \frac{\mathrm{d} \Delta \mathbf{v} }{\mathrm{d} t} = -\Omega^{2} z \displaystyle\frac{\left(\rhogz + \rhodz \cos \left(\Omega t \right) \right)}{\cos \left(\Omega t \right)\left( \rhogz \cos \left(\Omega t \right) + \rhodz \right)} \right\rbrace  & = - \Omega^{2} \zm \displaystyle \frac{\rhogz}{\rhodz} \label{eq:sys5} ,\\
\lim\limits_{\substack{\Omega t \to \pi / 2^{-}}}  \left\lbrace - (\deltav \cdot \nabla) \vb + \frac{1}{2}\nabla \left( \frac{\rhod - \rhog}{\rhod + \rhog} \deltav ^{2} \right) =  - \Omega^{2}\zm \frac{\sin\left(\Omega t \right)^{2}}{\cos\left(\Omega t \right)} + \left( \frac{\displaystyle \frac{\rhodz}{\cos\left(\Omega t \right)} - \rhogz}{\displaystyle \frac{\rhodz}{\cos\left(\Omega t \right)} + \rhogz} \right)\Omega^{2} z \tan \left(\Omega t \right)^{2} \right\rbrace &  = - \Omega^{2} \zm \displaystyle \frac{\rhogz}{\rhodz} \label{eq:sys6} .
\end{align}
%\end{multicols}
%
When dust particles reach $z = 0$, the density rate $\mathrm{d} \rho/\mathrm{d} t$ diverges (Eq.~\ref{eq:sys1}) while the total mass remains constant. Capturing this singularity represents a challenge for the numerical algorithm. By contrast, Eq.~\ref{eq:sys2} shows that the time derivative of the dust fraction remains finite. From Eq.~\ref{eq:sys3}, the acceleration of the one-fluid particle also does not diverge. Eq. ~\ref{eq:sys4} shows that its value when particles reach $z = 0$ is determined by the anisotropic pressure term, with the other terms being negligible. The \textsc{dustyoscill} problem is therefore useful for testing the implementation of the anisotropic pressure term, which it is negligible for most of the other problems we have considered. Even more noticeably, Eq.~\ref{eq:sys5} shows that the time derivative of the differential velocity $\mathrm{d} \Delta \mathbf{v} /\mathrm{d} t$ also remains finite. Eq.~\ref{eq:sys6} shows that this finite limit arises from the contribution of the two terms that are specific to the one-fluid mixture. Importantly, those terms diverge when taken separately. Thus, to deal with the \textsc{dustyoscill} problem accurately, the numerical algorithm has to integrate those term properly, so that they counterbalance each other effectively and lead to a finite value for $\mathrm{d} \Delta \mathbf{v} /\mathrm{d} t$ when particles cross $z = 0$.

\section{Implicit integration with non-linear drag coefficients}
\label{app:impquad}

In this Appendix, we detail the calculations involved in the implicit scheme for a quadratic non-linear drag regime, where $\deltav$ satisfies
\begin{equation}
\frac{\mathrm{d} \deltav}{\mathrm {d}t} = - \frac{K ' \left| \deltav \right|}{\rho \epsilon \left(1 - \epsilon \right)} \deltav+ a_{\rm 0} .
\label{eq:drag_NL_gene}
\end{equation}
Here, both the drag coefficient $K$ and the stopping time $\ts$ depend on $\deltav$ according to $K = K' \left| \deltav \right| $ and $\ts = \rho \epsilon \left(1 - \epsilon \right) / K\left(\deltav \right)$. To solve Eq.~\ref{eq:drag_NL_gene}, we rewrite it in the more convenient form
\begin{equation}
\frac{\mathrm{d} \deltav}{\mathrm {d}t} = - \frac{\left| \deltav \right| \deltav}{\ld} + a_{\rm 0},
\label{eq:drag_NL}
\end{equation}
where $\ld$ is the drag \emph{length}. We define $\vdt \equiv \sqrt{\az \ld}$ and $\td \equiv \ld / \vdt = \sqrt{\ld / \az}$ as the (magnitude of the) terminal velocity of the grain and the physical drag time, respectively. Eq.~\ref{eq:drag_NL} consists of two Ricatti equations (one for each sign of the absolute value). The solution of Eq.~\ref{eq:drag_NL} is given by
\begin{equation}
\deltav \left( t \right) = 
\begin{cases}
\dst \siga \vdt \dst \frac{ \dst \tanh\left( \frac{t}{\td}\right) +  \frac{\left|\deltavz\right|}{\vdt} }{ 1 + \dst \tanh\left( \frac{t}{\td}\right) \frac{\left|\deltavz\right|}{\vdt}  }   &; \siga \sigv = 1 ,\\
\begin{cases}
\dst \siga \vdt \dst \frac{ \dst \tan\left( \frac{t}{\td}\right) -  \frac{\left|\deltavz\right|}{\vdt} }{ 1 + \dst \tan\left( \frac{t}{\td}\right) \frac{\left|\deltavz\right|}{\vdt}  }, & \dst t \le t_{\rm c}  ,\\
\dst \siga \vdt \tanh\left( \frac{t - t_{\rm c}}{\td} \right), &t > t_{\rm c}
\end{cases}
    & ; \siga \sigv = -1 ,\\
  \siga \vdt \tanh \left(\frac{t}{\td} \right)  & ; v_{\rm 0} = 0,\\
\dst \frac{\deltavz}{1 + \dst \sigv \frac{\deltavz t}{\ld} }    &;  a_{\rm 0} = 0,
\end{cases}
\label{eq:sol_NL}
\end{equation}
where $\siga \equiv \mathrm{sgn}\left(a_{\rm 0} \right)$, $\sigv \equiv \mathrm{sgn}\left( \deltavz \right)$, $\deltavz = \deltav(t = 0)$ and
\begin{equation}
t_{\rm c} \equiv \td \arctan \left(\frac{\left| \deltavz \right|}{\vdt}\right) .
\label{eq:def_tc}
\end{equation}
The last case ($a_{\rm 0}$ = 0) is simply the solution of the \textsc{dustybox} problem with a quadratic drag regime derived in \citet{LP11}. The integral 
\begin{equation}
\mathcal{D} = \int_{0}^{\Delta t} \deltav\left(t' \right)^{2} \mathrm{d}t' 
\label{eq:}
\end{equation}
involved in the calculation of the internal energy term is given by
\begin{equation}
\mathcal{D} = 
\begin{cases}
\dst  \az \ld \left\lbrace  \Delta t + \td \left( \frac{\left| \deltavz \right|}{\vdt}   - \dst \frac{ \dst \tanh\left( \frac{t}{\td}\right) +  \frac{\left|\deltavz\right|}{\vdt} }{ 1 + \dst \tanh\left( \frac{t}{\td}\right) \frac{\left|\deltavz\right|}{\vdt}  }\right) \right\rbrace  &; \siga \sigv = 1 ,\\
\begin{cases}
\dst \tilde{\mathcal{D}}\left(\Delta t \right) = \az \ld \left\lbrace  - \Delta t + \td \left( \frac{\left| \deltavz \right|}{\vdt}   + \dst \frac{ \dst \tan\left( \frac{t}{\td}\right) -  \frac{\left|\deltavz\right|}{\vdt} }{ 1 + \dst \tan\left( \frac{t}{\td}\right) \frac{\left|\deltavz\right|}{\vdt}  }    \right) \right\rbrace &  ,  \dst \Delta t \le t_{\rm c}  ,\\
\dst \tilde{\mathcal{D}}\left(\Delta t_{\rm c} \right) + \Delta t - t_{\rm c} - \td \tanh\left(\frac{\Delta t - t_{\rm c}}{\td} \right), &t > t_{\rm c}
\end{cases}
    &  ; \siga \sigv = -1 ,\\
\dst  \az \ld \left\lbrace  \Delta t - \td \tanh\left( \frac{\Delta t}{\td} \right)\right\rbrace   & ; v_{\rm 0} = 0,\\
\dst \frac{\deltavz^{2}}{1 + \dst  \sigv\frac{ \deltavz \Delta t}{\ld}}   \Delta t  &;  a_{\rm 0} = 0.
\end{cases}
\label{eq:sol_NLu}
\end{equation}
Fig.~\ref{fig:NL_quad} shows that the accuracy of the implicit scheme in this non-linear drag regime is the same as the one obtained for a linear drag regime. Note that the algorithm is designed for a 1D integration. For a 3D problem, solutions of Matrix Ricatti equations should replace the solutions of the one-dimensional Ricatti equations used to solve Eq.~\ref{eq:drag_NL}.
\end{onecolumn}
\begin{twocolumn}
\begin{figure}
\begin{center}
   \includegraphics[width=\columnwidth]{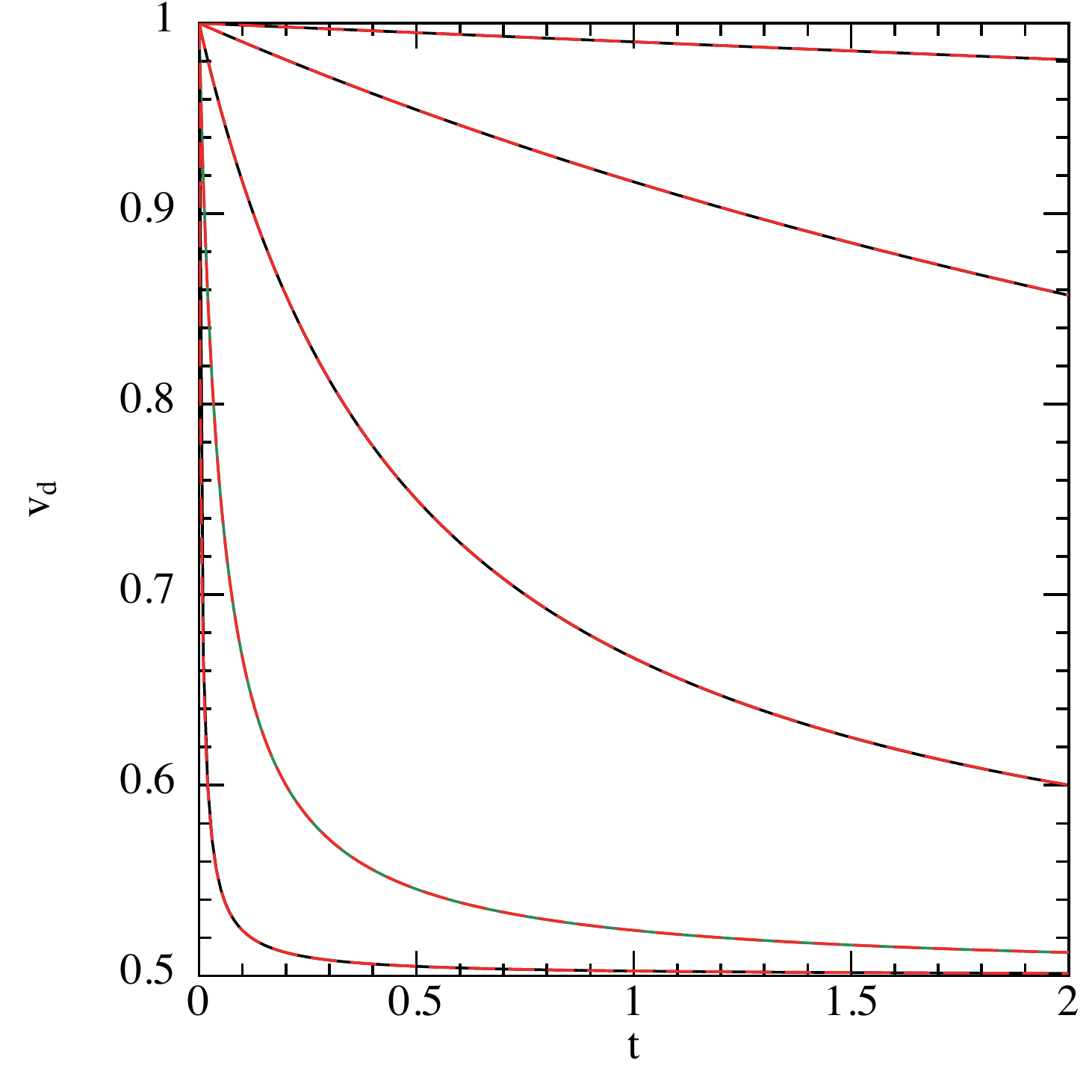}
   \caption{Simulations of the \textsc{dustybox} test with a quadratic non-linear drag regime. Otherwise the same as Fig.~\ref{fig:K}. As for the linear case, results are in excellent agreement with the exact solution given by the long-dashed/red lines ($< 1\%$ error in $L_{1}$).}
\end{center}
\end{figure}

\begin{figure}
\begin{center}
   \includegraphics[width=0.8\columnwidth]{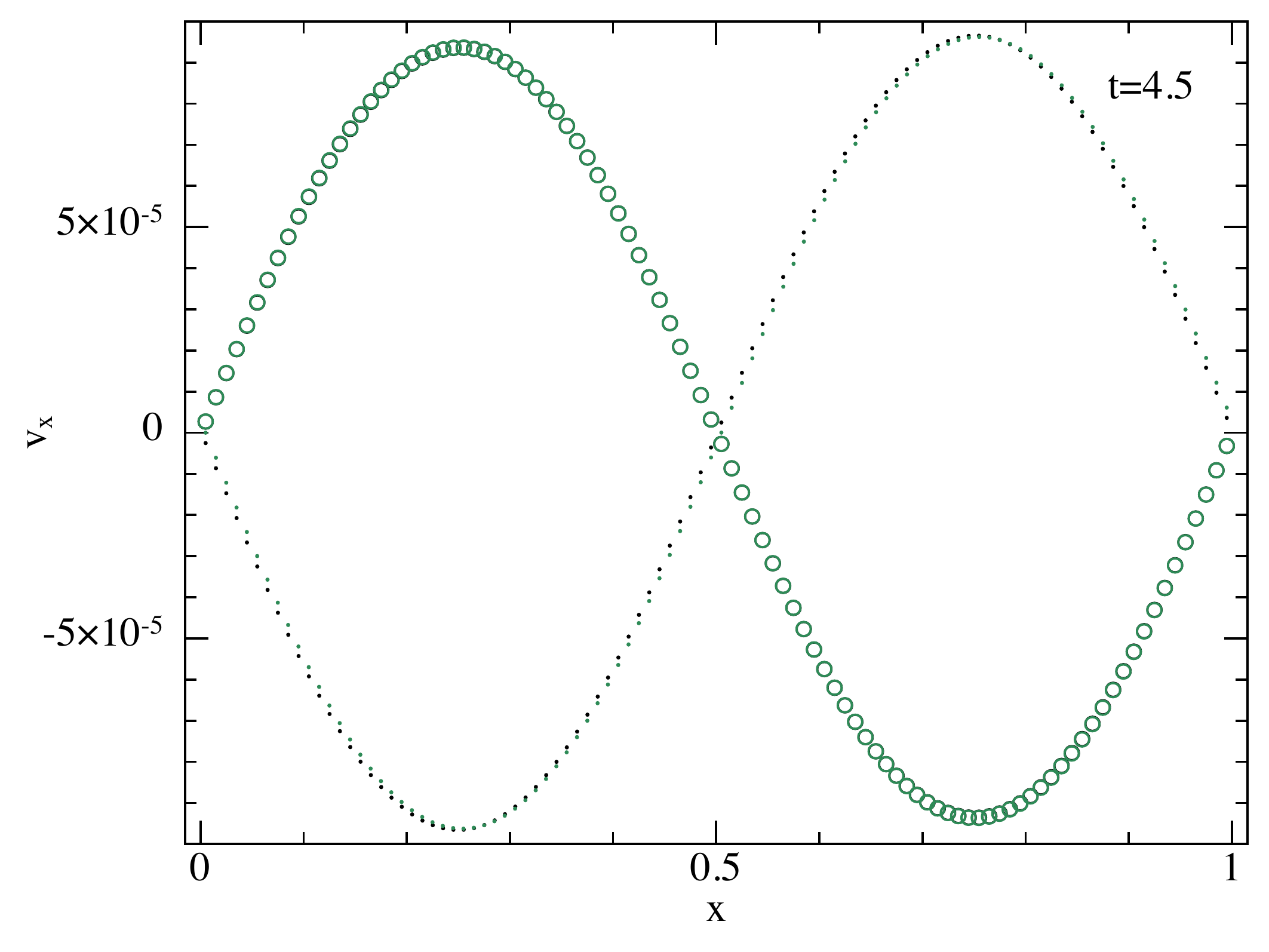} 
   \caption{Results of the \textsc{dustywave} test using a non-linear (quadratic) drag coefficient. As in Fig.~\ref{fig:dustywave} with $K = 100$ and a quadratic dependence on $\deltav$. Results are in excellent agreement with the results obtained by explicit integration with the two-fluid algorithm.}
   \label{fig:NL_quad}
\end{center}
\end{figure}
\end{twocolumn}
\end{appendix}

\label{lastpage}
\end{document}